\documentclass[journal]{IEEEtran}
\usepackage{graphicx}
\usepackage{hyperref}
\usepackage{xcolor}
\usepackage{color}
\usepackage{amsmath}
\usepackage{amsfonts}
\usepackage{amssymb}
\usepackage{esint}

\usepackage{multirow}

\newcommand{\ie}{\textit{i.e.}, }
\newcommand{\eg}{\textit{e.g.}, }

\newcommand{\ree}[1]{\textcolor{black}{#1}}

\ifCLASSINFOpdf
\else
\fi
\begin{document}
%
\title{\LARGE \bf Online Learning Koopman operator for closed-loop electrical neurostimulation in epilepsy}
%
%
%

\author{Zhichao Liang$^{\dagger}$, Zixiang Luo$^{\dagger}$, Keyin Liu, Jingwei Qiu  and Quanying Liu$^{*}$
\thanks{This work was funded in part by the National Key Research and Development Program of China (2021YFF1200800), National Natural Science Foundation of China (62001205),  Shenzhen-Hong Kong-Macao Science and Technology Innovation Project (SGDX2020110309280100), Guangdong Natural Science Foundation Joint Fund (2019A1515111038), Shenzhen Science and Technology Innovation Committee (20200925155957004, KCXFZ2020122117340001), Shenzhen Key Laboratory of Smart Healthcare Engineering (ZDSYS20200811144003009).}
\thanks{Z. Liang, Z. Luo, K. Liu, J. Qiu and Q. Liu are with Shenzhen Key Laboratory of Smart Healthcare Engineering, Department of Biomedical Engineering, Southern University of Science and Technology, Shenzhen, 518055, P. R. China.}
\thanks{${\dagger}$ Co-first authors: Zhichao Liang, Zixiang Luo}
\thanks{* Corresponding author: Quanying Liu {\tt\small liuqy@sustech.edu.cn}}%
}

\maketitle

\begin{abstract}
Electrical neuromodulation as a palliative treatment has been increasingly used in the control of epilepsy. However, current neuromodulations commonly implement predetermined actuation strategies and lack the capability of self-adaptively adjusting stimulation inputs. In this work, rooted in optimal control theory, we propose a \textit{Koopman-MPC framework} for real-time closed-loop electrical neuromodulation in epilepsy, which integrates i) a deep Koopman operator based dynamical model to predict the temporal evolution of epileptic EEG with an approximate finite-dimensional linear dynamics and ii) a model predictive control (MPC) module to design optimal seizure suppression strategies. 
The Koopman operator based linear dynamical model is embedded in the latent state space of the autoencoder neural network, in which we can approximate and update the Koopman operator online. The linear dynamical property of the Koopman operator ensures the convexity of the optimization problem for subsequent MPC control. 
The proposed deep Koopman operator model shows greater predictive capability than the baseline models (\eg vector autoregressive model, kernel based method and recurrent neural network (RNN)) in both synthetic and real epileptic EEG data.
Moreover, compared with the RNN-MPC framework, our Koopman-MPC framework can suppress seizure dynamics with better computational efficiency in both the Jansen-Rit model and the Epileptor model.
Koopman-MPC framework opens a new window for model-based closed-loop neuromodulation and sheds light on nonlinear neurodynamics and feedback control policies.
\end{abstract}


%
\IEEEpeerreviewmaketitle

\IEEEPARstart{E}{pilepsy} is a neurological disorder characterized by the occurrence of spontaneous epileptic seizures, during which the neuronal population fires in an abnormal, excessive and synchronized manner~\cite{fisher2014ilae,yuan2021machine}. \ree{
Clinically, the main biomarkers for the diagnosis of epilepsy with electrophysiological signals are epileptiform spikes and seizure-like waves, which are usually monitored by noninvasive electroencephalogram (EEG). The invasive stereoelectroencephalogram (sEEG) and intracranial
electroencephalogram (iEEG) are utilized to better localize the seizure onset zone~\cite{wetjen2009intracranial,jayakar2016diagnostic,cogan2017multi}.}

A range of treatments have been developed to control seizures, such as anti-epilepsy medication, palliative/radical surgery, and electrical neurostimulation~\cite{yuan2021machine,lowe2004epilepsy,moshe2015epilepsy,berenyi2012closed,salam2015seizure}. Compared \ree{to surgical resections}, electrical neurostimulation, \eg deep brain stimulation (DBS) and transcranial electrical stimulation (TES), has fewer long-term side effects on brain function, and has been becoming a considerable clinical intervention for intractable epilepsy~\cite{berenyi2012closed,salanova2018deep,li2018deep}. \ree{For instance, electrical nerve stimulation applies a certain dose of high-frequency current within a safe range} (\eg $50~Hz$) to a specific targeting region (i.e., $0.5\sim5~mA$ in the mesial temporal cortex), aiming at modulating the seizure-like neural oscillations to the seizure-free state~\cite{ berenyi2012closed,oderiz2019association}.

However, current neuromodulation mainly implements predetermined open-loop actuation strategies, rather than \ree{a closed-loop stimulation strategy that is updated in real time}. 
The open-loop neurostimulation actually lacks feedback signals from the observed real-time seizure waves for adjusting the control inputs during neurostimulation, which would limit its flexibility in practical applications~\cite{salam2015seizure}. More recently, closed-loop neuromodulation is increasingly gaining attention in clinical applications such as seizure control~\cite{berenyi2012closed,salam2015seizure,scangos2021closed}. 
By framing the neuromodulation as a closed-loop state-space control problem, the main goal then is to design an optimal control law to steer the present system state to the desired state. Some pioneering works have been devoted to this area, ranging from designing state-space models to designing control methods. For example, Pequito et al. presented a spectral control method to ensure that the poles of a closed-loop system are within the prespecified spectrum range~\cite{2017Spectral}. 
Ashourvan et al. proposed a pole-placement spectral static output feedback control-theoretic strategy for suppressing seizure dynamics~\cite{2020Model}. 
Recently, model predictive control (MPC), as a model-based and self-tuning control method to maintain a stable and robust control process, has also been brought into neurostimulation applications~\cite{chang2020model,chatterjee2020fractional}. For instance, Chang et al. integrated a nonlinear auto-regressive moving-average (NARMA) Volterra model as an identified dynamical system into MPC for seizure suppression~\cite{chang2020model}. Chatterjee et al. proposed a fractional-order model based MPC framework in neuromodulation~\cite{chatterjee2020fractional}.

Seizure dynamics is a complex networked nonlinear dynamic process, which we have not yet fully understood~\cite{jirsa2014epileptor,verdugo2019glia,arrais2021design}. The control of seizures requires modeling the underlying complex seizure dynamics and obtaining optimal control law for seizure suppression, as opposed to the seizure detection or classification tasks with deep learning solutions~\cite{li2021spatio,zhang2019epilepsy,zabihi2019patient,alam2013detection,guo2018stacked,craley2019spatio,zhang2016design}. Therefore, system identification is crucial for uncovering the system dynamics over the seizure period by building a dynamical model and estimating 
its parameters based on the observed input-output data. 
Several traditional system identification methods have been used to model the seizure dynamics with EEG data, such as the auto-regressive moving-average model~\cite{chang2020model} and the fractional-order system model~\cite{chatterjee2020fractional}. However, \ree{these two methods have not yet been validated in suppressing network-coupled seizure dynamics.} With advances in machine learning, data-driven models with neural networks have shown great potential in system identification~\cite{hazan2017learning,lanzetti2019recurrent,bieker2020deep}. For instance, recurrent neural networks (RNNs) have proven promising in fitting time-series data and modeling dynamical systems in fluid flow control~\cite{bieker2020deep} and process industries~\cite{lanzetti2019recurrent}. 
However, \ree{integrated the highly nonlinear RNN models into a MPC framework for real-time optimization control, the quadratic programming problem becomes a non-convex optimization problem that is computationally inefficient. }



When modeling seizure dynamics and designing seizure suppression controllers in the MPC framework, it is necessary to consider the trade-off between model accuracy and complexity. Higher model accuracy improves the predictive performance for approximating the nonlinear seizure system, while lower model complexity guarantees computational efficiency when performing MPC optimizations.
\ree{For the non-convex nonlinear quadratic programming optimization problem originated from MPC optimization of a nonlinear complex dynamic model, finding its optimal solution is theoretically difficult and numerically inefficient.}
\ree{As an alternative,} operator-theoretic approaches, mainly based on the Perron-Frobenius operator~\cite{Andrzej1994Chaos} or its adjoint Koopman operator~\cite{Koopman1931hamiltonian}, \ree{aim to approximate} a nonlinear dynamical system with linear operators. \ree{These operator based methods can help avoid the non-convex MPC optimization problems.}
However, the predefined Koopman operator model has substantial limitations in modeling time-varying or complex switching dynamical systems. Therefore, a more flexible way to approximate the finite-dimensional Koopman operator in an invariant subspace is needed~\cite{takeishi2017learning}. Recently, autoencoder architecture has been employed to learn the finite-dimensional Koopman operator instead of the hand-crafted design of the basis function \ree{in the prediction of fluid flow dynamics}~\cite{morton2018deep,takeishi2017learning}. \ree{Considering the linear properties of the Koopman operator and its successful applications in predicting fluid flow dynamics (complex nonlinear system), we aim to integrate the Koopman operator with model predictive control for closed-loop neurostimulation.} 

In this study, we propose a deep Koopman operator based model predictive control framework (\ie Koopman-MPC framework) for closed-loop seizure suppression in a real-time manner. 
The Koopman-MPC framework can meet the needs of sufficient prediction accuracy and low model complexity for real-time optimization.
The major contributions of this study are summarized as follows.
\begin{itemize}
  \item A Koopman-MPC framework is proposed (\textbf{Sec.~\ref{sec:method}}), including a \textit{Koopman operator based linear dynamical model} embedded into a finite-dimensional invariant subspace, and a \textit{MPC controller} to optimize the neurostimulation policy for seizure control. The design of Koopman-based model leverages the flexibility of online learning of Koopman operator, and enables real-time optimization for MPC controller.
  \item A deep autoencoder neural network is employed to learn, rather than to hand-crafted design, the embedding linear invariant subspace (\textbf{Sec.~\ref{sec:Koopman}}). Experimental results demonstrate that the deep Koopman operator model outperforms other baseline models in seizure prediction, for both synthetic data and real iEEG data (\textbf{Sec.~\ref{sec:results-Prediction}}). 
  \item The seizure suppression performance of the Koopman-MPC framework is validated and compared with the baseline RNN-MPC framework in \ree{both} Jansen-Rit model \ree{and Epileptor model}. The results show that Koopman-MPC framework successfully suppresses seizure dynamics with higher computational efficiency (\textbf{Sec.~\ref{sec:results-MPC}}) and allows online learning of the Koopman operator for closed-loop epilepsy control (\textbf{Sec.\ref{online_learning_seizure_suppression}}). 
\end{itemize}
\section{Koopman-MPC framework}
\label{sec:method}
In this section, we first introduce the Koopman operator theory and its extension to the linear forced dynamical systems, and then we propose the Koopman operator based model predictive control for obtaining the optimal control signals over the Koopman observations.

\subsection{Koopman operator}
\label{koopman_theory_section}
Consider a nonlinear discrete-time dynamical system
\begin{equation} \label{Dynamics}
    \emph{\textbf{x}}_{i+1}=\emph{f}(\emph{\textbf{x}}_i),
\end{equation}
where $ \emph{\textbf{x}}_{i} \in \mathcal{M} \subset \mathbb{R}^{n} $ denotes the vector of state variables at time step $i$ and the function $f(\cdot):\mathcal{M} \to \mathcal{M}$ governs nonlinear dynamics of the system under a finite low-dimensional manifold $\mathcal{M}$, in which the future state $\emph{\textbf{x}}_{i+1}$ only depends on the current state $\emph{\textbf{x}}_{i}$.

\begin{figure}
\centering
\includegraphics[width=0.9\linewidth]{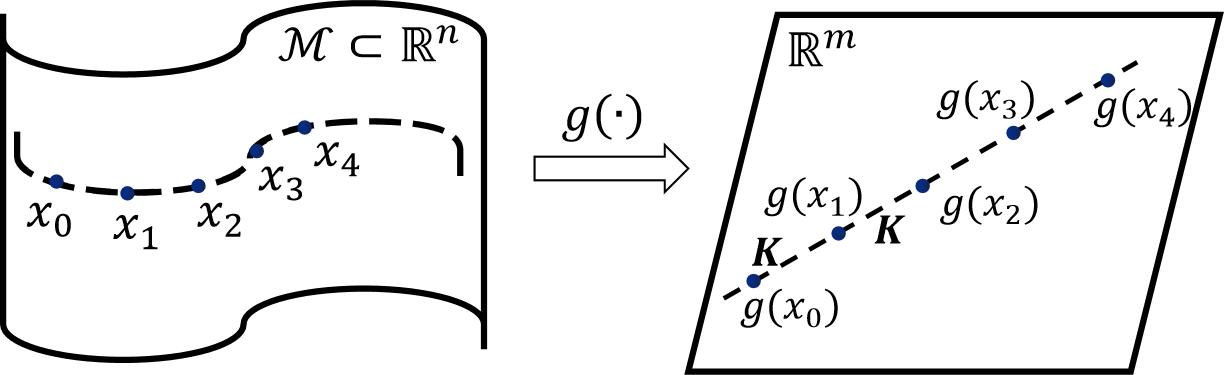}
\caption{Koopman operator theory. $g(\cdot)$ is a transformation from a nonlinear manifold space $\mathcal{M}\subset \mathbb{R}^{n}$ to a linear manifold space $\mathbb{R}^{m}$ \ree{that satisfies the linear forward relation $K$.}}
\label{fig:koopman_theory}
\centering
\end{figure}

According to the Koopman operator theory, there exists an infinite-dimensional linear operator $\emph{\textbf{K}}$ defined on an infinite-dimensional function space such that the observation functions $g(\cdot)$ satisfy an advancing linear forward  relation, shown in Fig.~\ref{fig:koopman_theory}.
\begin{equation} \label{Koopman}
    g( \emph{\textbf{x}}_{i+1} ) = \emph{\textbf{K}} g(\textbf{x}_{i})= g\circ \emph{f}(\emph{\textbf{x}}_i).
\end{equation}

The modal decomposition of the Koopman operator \emph{\textbf{K}} refers to spectral analysis (Eigendecomposition)~\cite{korda2020data}. 
\ree{The eigenvalue and corresponding eigenvector of Koopman operator \emph{\textbf{K}} have specific physical meanings.} The eigenvalues of \emph{\textbf{K}} reflect system stability, exponential damping ratio and spectral information, while the eigenvectors represent the independent spatial coordinates in the Koopman invariant space.


One nonnegligible issue about the Koopman operator modeling is how to design (or learn) the nonlinear embedding function $g(\cdot)$. 
A variety of methods have been proposed to design the finite-dimensional embedding function $\hat{g} (\cdot)$ to approximate the infinite-dimensional space (\textit{i.e.,} the extended dynamic mode decomposition~\cite{williams2015data} and auto-encoder~\cite{morton2018deep,takeishi2017learning}). 


Given a finite sequence of system state measurements $\emph{\textbf{x}}_i$ ($i = \{0, 1,\dots, n\}$), 
\ree{the extended dynamic mode decomposition method constitutes a basis function vector $\Phi = [\Phi_1,\Phi_2,...,\Phi_{m}]^T$, }
to construct two feature states $\Psi_{X}$ and $\Psi_{\hat{X}}$, as follows:
\begin{equation}
    \label{Feature Space}
    \begin{array}{cc}
         & \Psi_{X} = [\Phi(\emph{\textbf{x}}_0)\quad\Phi(\emph{\textbf{x}}_1)\quad  ...  \quad\Phi(\emph{\textbf{x}}_{n-1})] \in \mathbb{R}^{m \times n}\\
         & \Psi_{\hat{X}} = [\Phi(\emph{\textbf{x}}_1)\quad\Phi(\emph{\textbf{x}}_2)\quad  ...  \quad\Phi(\emph{\textbf{x}}_{n})] \in \mathbb{R}^{m \times n}
    \end{array}
\end{equation}

According to the linear least-square estimation, the finite-dimensional approximation of the Koopman operator $\hat{\emph{\textbf{K}}} \in \mathbb{R}^{m\times m} $ can be inferred as
\begin{equation}
    \label{Koopman Operator}
    \hat{\emph{\textbf{K}}} = \Psi_{\hat{X}}\Psi_{X}^{\dagger}
\end{equation}
where $\Psi_{X}^{\dagger}$ is the pseudo inverse of $\Psi_{X}$ and defined as $\Psi_{X}^{\dagger} = \Psi_{X}^T(\Psi_{X}\Psi_{X}^T+\lambda \textbf{I})^{-1}$, and $\lambda$ is the regularization factor. 

The approximated linear dynamical system is rewritten as
\begin{equation}
    \label{Koopman_Linear}
    \Phi(\emph{\textbf{x}}_{i+1})=\hat{\emph{\textbf{K}}} \Phi (\emph{\textbf{x}}_{i}).
\end{equation}

Let us consider a nonlinear forced dynamics with input $\textbf{u}_i$,
\begin{equation} \label{Dynamics}
    \emph{\textbf{x}}_{i+1}=\emph{f}(\emph{\textbf{x}}_i, \textbf{u}_i),
\end{equation}
where $\textbf{u}_i \in \mathbb{R}^{r \times 1}$ is a $r$-dimension control input at time step $i$, namely there are $r$ inputs in this system. 

Then, we extend the approximated Koopman operator based linear dynamical system into a linear forced dynamical system, described as Eq.~\eqref{Koopman_forced}.
In this case, the objective is simultaneously to learn an approximated finite-dimensional representation space of the state variables and its relationship with control inputs. 
\begin{equation}
    \label{Koopman_forced}
    \begin{aligned}
        \Phi(\emph{\textbf{x}}_{i+1})&=\hat{\emph{\textbf{K}}} \Phi (\emph{\textbf{x}}_{i})+\hat{\emph{\textbf{B}}}\textbf{u}_i \\
        & =[\hat{\emph{\textbf{K}}},\hat{\emph{\textbf{B}}}] \left[\begin{array}{l} \Phi (\emph{\textbf{x}}_{i}) \\ \textbf{u}_i \end{array}\right].
    \end{aligned}
\end{equation}
where $\hat{\emph{\textbf{B}}} \in \mathbb{R}^{m \times r} $ is the control matrix. Then, the approximated Koopman operator and the control matrix can be inferred from
\begin{equation}
    \label{Koopman Operator with forced}
    [\hat{\emph{\textbf{K}}},\hat{\emph{\textbf{B}}}] = \Psi_{\hat{X}} \hat{\Psi}_{X}^{\dagger},
\end{equation}
where $\hat{\Psi}_{X}=\left[\begin{array}{l} \Psi_{X} \\ U \end{array}\right]
               =\left[\begin{array}{llll}
                \Phi(\emph{\textbf{x}}_1) & \Phi(\emph{\textbf{x}}_2) &\ldots &\Phi(\emph{\textbf{x}}_n)  \\
                     \textbf{u}_1 & \textbf{u}_2 &\ldots & \textbf{u}_n
                \end{array}\right]$ is the augmented state.

In this study, we design a deep autoencoder neural network as the basis function to learn the finite-dimensional linear invariant subspace in a self-supervised manner (Sec.~\ref{sec:Koopman}) instead of the hand-crafted basis function ${\Phi}(\cdot)$.

\subsection{Koopman-MPC Framework for Seizure Suppression}
Model predictive control is an optimization-based control framework. It minimizes the objective function in a finite prediction horizon with the control inputs and states transition constraints. The MPC optimization for highly nonlinear dynamical systems is a nonconvex problem of NP-hardness, and we prefer to \ree{model it as the MPC convex optimization problem on linear systems with a unique solution.} The predictive capability and linearity of the approximated Koopman operator facilitate the design of the model predictive controller. 


\begin{figure}[t]
\centering
\includegraphics[width=\linewidth]{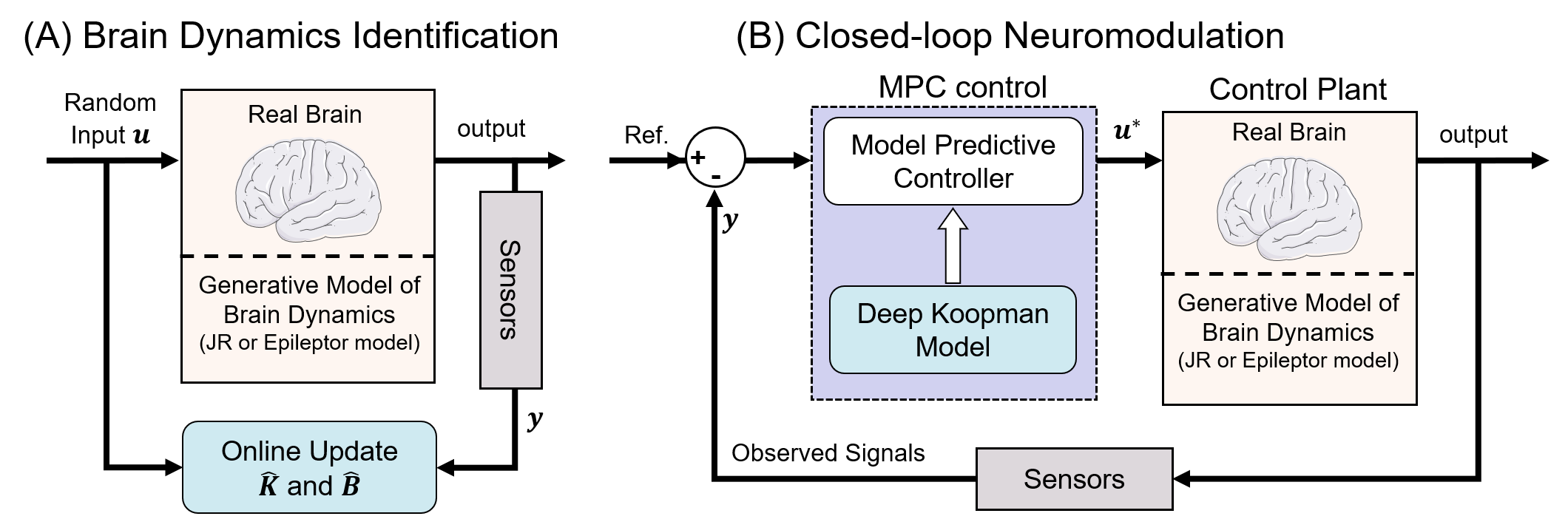}
\caption{\ree{The Koopman operator based MPC closed-loop neurostimulation framework for seizure suppression. The control plant can be the real brain or a generative model of brain dynamics (\eg the Jansen-Rit model or the Epileptor model). In the brain dynamics identification (A), the system parameters (\ie Koopman operator $\hat{K}$ and Control matrix $\hat{B}$) are identified by the input-output data pair ($u,y$). In the closed-loop neuromodulation (B), the optimal control $u^*$ is solved by MPC with the identified deep Koopman model.}}
\label{fig:framework}
\centering
\end{figure}

In this study, we construct the Koopman-MPC framework to control the nonlinear dynamics of seizures following the desired reference signals (seizure-free signals), as shown in Fig. \ref{fig:framework}. 
\ree{The objective function is defined as follows,}
\begin{equation}
\label{Eq:MPC}
    \begin{array}{cl}
\min\limits_{\Delta \textbf{u}} &\sum\limits_{i=1}^{T_p}\left(\Phi(\emph{\textbf{x}}_{i})-\Phi(\emph{\textbf{x}}_{ref})\right)^{T} \textbf{Q}_{X}\left(\Phi(\emph{\textbf{x}}_{i})-\Phi(\emph{\textbf{x}}_{ref})\right) \\
&+\sum\limits_{i=1}^{T_c}\Delta \textbf{u}_{i}^{T} \textbf{Q}_{u} \Delta \textbf{u}_{i} \\
\text{ s.t. }  & \Phi(\emph{\textbf{x}}_{i+1}) = \hat{\emph{\textbf{K}}}\Phi(\emph{\textbf{x}}_{i})+\hat{\emph{\textbf{B}}}\textbf{u}_{i} \\
& \textbf{u}_i =\textbf{u}_{i-1}+\Delta \textbf{u}_{i}  \\
& \textbf{u}_i \in\left[\textbf{u}_{\min}, \textbf{u}_{\max}\right]  \\
& \Delta \textbf{u}_{i} \in\left[\Delta \textbf{u}_{\min}, \Delta \textbf{u}_{\max}\right] 
\end{array}
\end{equation}
where $T_p$ and $T_c$ are the length of predictive horizon and control horizon,  respectively. $\Phi(\emph{\textbf{x}}_{i})$ and $\Phi(\emph{\textbf{x}}_{ref}) \in \mathbb{R}^{m \times 1}$ are the finite-dimensional representation of the system state $\emph{\textbf{x}}_{i}$ and reference signals $\emph{\textbf{x}}_{ref}$ in the Koopman invariant subspace, respectively. $\Delta \textbf{u}_{i} \in \mathbb{R}^{r \times 1}$ ($r$ is the number of system inputs) is the increment of input at step ${i}$. $\textbf{Q}_{X} \in \mathbb{R}^{m \times m}$ is the positive definite weighted matrix for penalizing the deviance and $\textbf{Q}_{u} \in \mathbb{R}^{r \times r}$ is a non-negative matrix for penalizing the incremental amplitudes of control inputs. For real applications, the inputs are constrained between the upper and lower bound. The equation $\Phi(\emph{\textbf{x}}_{i+1}) = \hat{\emph{\textbf{K}}}\Phi(\emph{\textbf{x}}_{i})+\hat{\emph{\textbf{B}}}\textbf{u}_{i}$ is the linear state transition constraint condition that predicts the future states. 
In this study, \ree{the input signals are constrained from -30 mV to 5 mV in the Jansen-Rit model and -4 mV to 0.5 mV in the Epileptor model. The increment of input signals is constrained from -20 mV to 0.5 mV in the Jansen-Rit model and -0.25mV to 0.05mV in the Epileptor model.} $\textbf{Q}_{X}$ is set to be an identity matrix $\mathbb{I}$. $\textbf{Q}_{u}$ equals to 0.01$\mathbb{I}$.

\ree{To solve the optimization problem in the Koopman-MPC framework in Eq~\eqref{Eq:MPC}, we utilize linear quadratic programming solver at each time step to find a sequence of incremental inputs $\Delta \textbf{u}_{i}$ $(i=1,2,\cdots,T_c)$, and only the first incremental input $\Delta \textbf{u}_1$ is applied to the system for seizure suppression.}

\section{Deep Koopman operator Based Seizure dynamical model}
\label{sec:Koopman}
\begin{figure}[htbp]

\centering
\includegraphics[width=\linewidth]{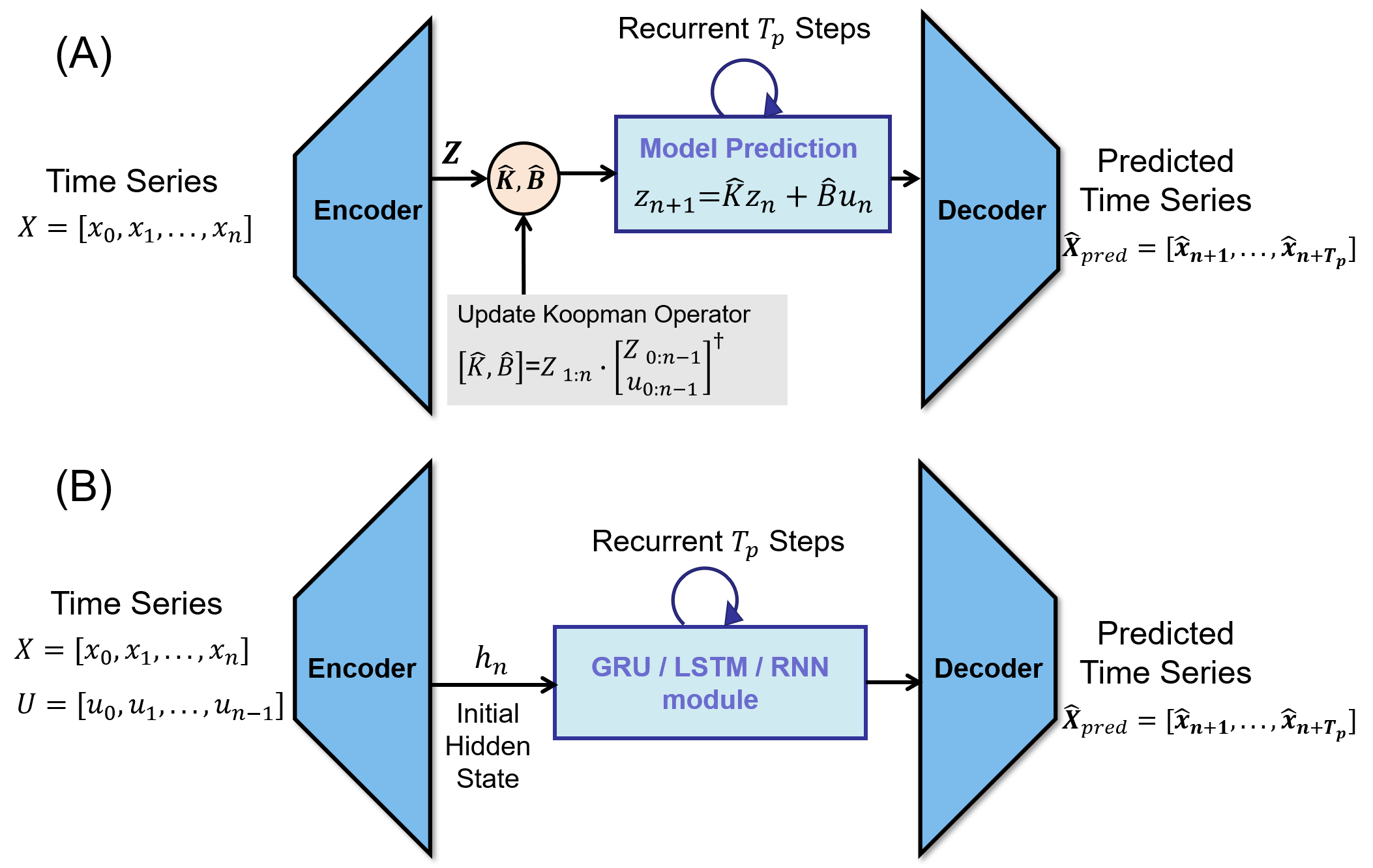}
\caption{\ree{The framework of (A) deep Koopman model and (B) RNN-type model. The encoder in the deep Koopman model transforms the original nonlinear manifold space into a linear forward state transition space. In contrast, the encoder in the RNN-type model infers the initial hidden state of the recurrent cell. Both of the decoders transform the states in hidden state space back into the original state space.}}
\label{fig:koopman_rnn_ae_framework}
\centering
\end{figure}

We design an autoencoder framework with fully-connected layers to learn a finite-dimensional invariant subspace, where we can approximate \ree{and} update the Koopman operator. The proposed deep Koopman operator model is shown in Fig.~\ref{fig:koopman_rnn_ae_framework} (A). \ree{As a comparison, the RNN-based predictive model is presented in }Fig.~\ref{fig:koopman_rnn_ae_framework} (B). \ree{Both the RNN-type models and Koopman-based model are composed of an encoder, a decoder and a dynamical system that predicts the future states} (Fig.~\ref{fig:koopman_rnn_ae_framework}). \ree{The main difference is that the dynamics in hidden state space of the RNN-type network is a nonlinear system, while the dynamics in hidden state space of the deep Koopman model is a linear system characterized by the Koopman operator. }


\subsection{Autoencoder Architecture to Learn the Linear Embedding of nonlinear Seizure Dynamics}
\label{AE_4_Learning_Seizure_Dynamics}
The Koopman operator based autoencoder consists of an encoder and a decoder neural network, both of which contain \ree{three independent fully-connected layers}. The first two layers are followed by a hyperbolic tangent activation function (tanh), while the last layer has no activation function and directly generates the output. 


\ree{As shown in Fig.\ref{fig:koopman_rnn_ae_framework} (A), $X$ is the input signal to the model, where $X=[x_0,x_1,\cdots,x_{n}]\in \mathbb{R}^{k \times (n+1)}$ denotes the past $n$ steps of data $x_{0:n-1}=[x_0,x_1,\cdots,x_{n-1}]$ and the current state $x_{n}$. $k$ is the dimension of the state space of the original system. 
The encoder embeds the low-dimensional input data into a high-dimensional latent space and generates $Z \in \mathbb{R}^{m \times (n+1)}$ ($m>k$),}
\begin{equation}
Z=encoder(X) 
\end{equation}

The following two cases are considered in the linear dynamical system within the finite-dimensional space.

(1) Without external inputs. 

In this case, the prediction process is governed by
\begin{equation}
\label{prediction_without_input}
     z_{n+1} = \hat{\emph{\textbf{K}}}z_{n}.
\end{equation}

\ree{We estimate $\hat{\emph{\textbf{K}}}$ by solving the linear least squares problem of 
$Z_{1:n} = \hat{\emph{\textbf{K}}} Z_{0:n-1}$.}


(2) With external inputs.

In this case, we consider the basic linear forced dynamical system for prediction
\begin{equation}
\label{prediction_with_input}
     z_{n+1} = \hat{\emph{\textbf{K}}}z_{n}+\hat{\emph{\textbf{B}}}u_{n}=[\hat{\emph{\textbf{K}}},\hat{\emph{\textbf{B}}}]  \left[\begin{array}{l} z_n \\ u_n \end{array}\right].
\end{equation}

\ree{We estimate the Koopman operator $\hat{\emph{\textbf{K}}}$ and control  matrix $\hat{\emph{\textbf{B}}}$ with $Z_{0:n-1}$, $Z_{1:n}$ and input series $u_{0:n-1}=[u_0,u_1,\cdots,u_{n-1}]$.}
\begin{equation}
    [\hat{\emph{\textbf{K}}},\hat{\emph{\textbf{B}}}] = Z_{1:n}  \left[\begin{array}{l} Z_{0:n-1} \\ u_{0:n-1} \end{array}\right]^{\dagger}
\end{equation}
\ree{where $\left[\begin{array}{l} Z_{0:n-1} \\ u_{0:n-1} \end{array}\right]^{\dagger}$ is the pseudo inverse of the augmented state of $Z_{0:n-1}$ and $u_{0:n-1}$.}

\ree{The one-step linear approximated Koopman dynamics can be extended to higher-order case, similar to the higher-order autoregressive model, by considering encoded inputs from more previous time steps $z_{n-i}  (i=0,1,\dots,r)$, where $r$ is the number of previous steps.}
\ree{
\begin{equation}
    \label{prediction_with_input_and_high_order}
     z_{n+1} = \sum_{i=0}^{r} \hat{\mathbf{K}}_{i} z_{n-i}+\hat{\mathbf{B}}u_{n}=[\hat{\mathbf{K}}_0,\cdots,\hat{\mathbf{K}}_{r},\hat{\mathbf{B}}]\left[\begin{array}{l} z_n \\ \dotsc \\ z_{n-r} \\ u_n \end{array}\right].
\end{equation}
}

\ree{The higher-order linear Koopman dynamics can still be solved by the linear least squares estimator.}

\ree{The seizure dynamic is predicted in the Koopman invariant subspace by rolling the linear forward equation as Eq.~\eqref{prediction_without_input},~\eqref{prediction_with_input} or ~\eqref{prediction_with_input_and_high_order} accordingly. }



The decoder enables us to transform $Z$ and \ree{the prediction of $z_{n+i}, (i=1,2,\cdots,T_p)$} back into the original state space
\begin{equation}
\hat{X}=decoder(Z), \  \hat{x}_{n+i}=decoder(z_{n+i}).
\end{equation}



\subsection{Definition of Explicit Loss Function}
The objective of the deep Koopman operator based autoencoder is to learn a finite invariant subspace that approximates the Koopman operator. The loss is specifically designed with different theoretical considerations for learning the finite-dimensional approximated Koopman operator and building the linear dynamical system. In summary, the explicit loss function for training the deep Koopman operator model covers the following reconstruction term and prediction-reconstruction term with Mean-Squared Error (MSE). 

(1) Reconstruction error of $X$:
\begin{equation}
\mathcal{L}_{\text{recon}}=\frac{1}{n*D}\left\|X-\tilde{X}\right\|^2_2,
\end{equation}

(2) Prediction-reconstruction error:
\begin{equation}
\mathcal{L}_{\text{pred}}=\frac{1}{T_p*D}\sum_{i=1}^{T_p}\left\|x_{n+i}-\tilde{x}_{n+i}\right\|^2_2,
\end{equation}
where $D$ is the dimension of the original state space. $n$ is the number of steps in both $X$ and $\tilde{X}$. $T_p$ is the length of prediction.


The total loss function can be summarized as
\begin{equation}
\mathcal{L}=\lambda_{recon}\mathcal{L}_{\text{recon }}+\lambda_{pred}\mathcal{L}_{\text {pred }}
\end{equation}
where $\lambda_{recon}$ and $\lambda_{pred}$ are the hyper-parameters to balance the reconstruction error and prediction-reconstruction error. In our study, we set them equal to 1 by default.

\section{Experiments and results}
We first test the performance of the deep Koopman operator model in predicting seizure dynamics using synthetic data and real data and then test the performance of the Koopman-MPC framework on seizure suppression using a simulation platform.

\subsection{Dataset Description}

\begin{figure}[htpb]
\centering
\includegraphics[width=\linewidth]{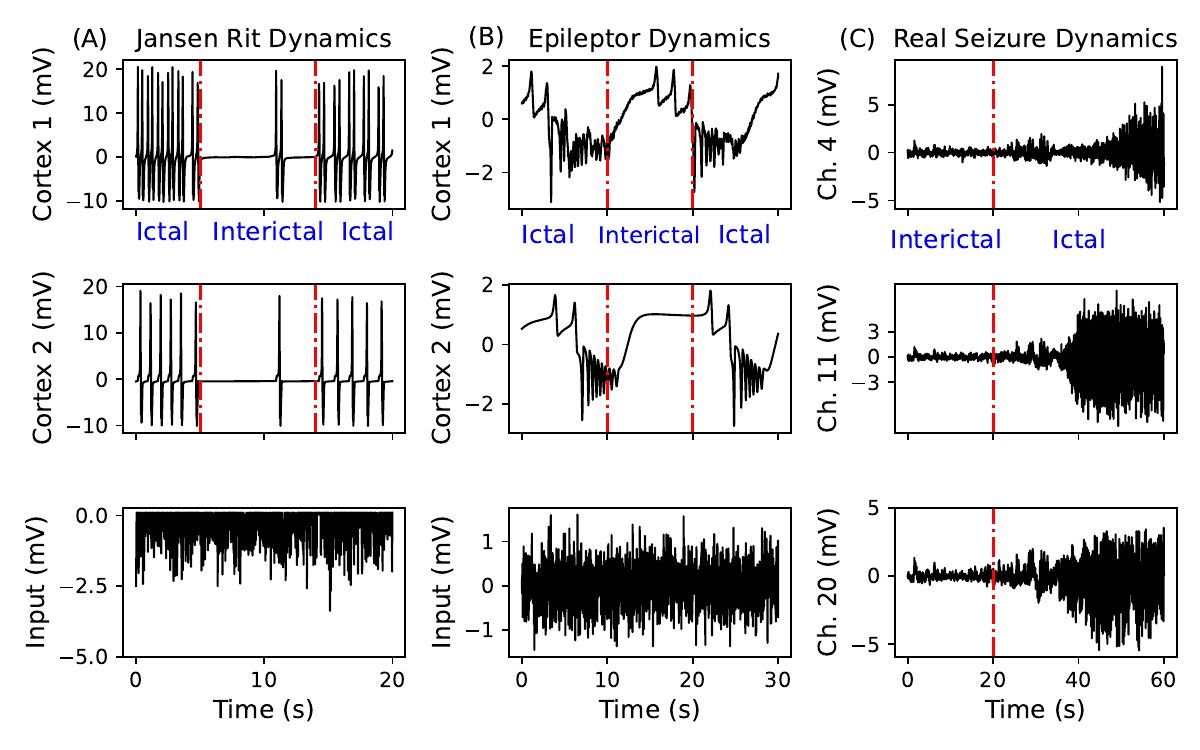}
\caption{\ree{The visualization of the synthetic data and real data. The synthetic seizure-like waves propagate from the epileptogenic foci (up) to a seizure-free cortex (bottom) in (A) the Coupled Jansen-Rit Model and (B) the Coupled Epileptor Model, (C) real data with state transition from interictal to ictal period.}}
\label{fig:JR-EPI-EEG}
\centering
\end{figure}

The deep Koopman operator model is validated with both synthetic data and real data.
We used the well-known \ree{Jansen-Rit model~\cite{jansen1995electroencephalogram} and the Epileptor~\cite{jirsa2014epileptor} model to generate seizure dynamics. By varying the model parameter (\ie the excitatory synaptic gain, $A$)}, Jansen-Rit model can synthesize the seizure-free EEG and seizure EEG signals. \ree{In our study, we mimic the seizure propagation process in two distant interconnected cortical columns by a coupled Jansen-Rit model described in~\cite{jansen1995electroencephalogram} and the seizure transition from interictal period to ictal period. The seizure propagates from the epileptogenic foci (Cortex 1) to a seizure-free region (Cortex 2). The seizure transition process is induced by altering the excitatory parameter $A$ in cortex 1. An example of synthetic data with seizure transition and seizure propagation is shown in Fig.~\ref{fig:JR-EPI-EEG}(A). The ordinary differential equations (ODEs) of the coupled Jansen-Rit model and its parameter settings are summarized in Appendix sec.~\ref{jansen_rit_model} and Table~\ref{table:para-double_JR-model}.} 

\ree{The Epileptor model integrates the dynamics in seizure interictal and ictal periods using ODEs with three different time scales, as described in Appendix sec.~\ref{epileptor_description}. Changing the excitability parameter $x_0$ in the Epileptor model can generate seizure oscillations or seizure-free dynamics. We also simulate the seizure propagation process in a coupled Epileptor model with two interconnected cortexs, as shown in Fig.~\ref{fig:JR-EPI-EEG}(B). } 

\ree{We synthesize 4000s recording with 100 Hz sampling rate in each synthetic model. The first half of each recording was separated as the training and another as testing data.}

\ree{The real data comes from an open dataset where the raw iEEG data for patients comes from NIH, UMH, UMMC, and JHH in the OpenNeuro repository (\url{https://openneuro.org/datasets/ds003029/versions/1.0.3}). It includes iEEG signals during the interictal period and ictal period with 1000 Hz sampling rate. The recording of the real data with randomly selected channels is shown in Fig.~\ref{fig:JR-EPI-EEG}(C). We select one of the patients with 84 channels and two different time stamp recordings. One of the recordings is used for training the model, and the other is used for testing.}



\subsection{Baseline models and evaluation metrics}

\ree{To evaluate the performance of data prediction, the Koopman operator model is compared with several baseline models, including two vector autoregressive models (\ie a 5-order VAR, a 10-order VAR), a vector autoregressive moving average model (\ie VARMA with 5-order AR and 5-order MA), a kernel-based model (\ie SINDy~\cite{brunton2016discovering}) and three types of recurrent neural network models (\ie Vanilla RNN, Gated Recurrent Units, and Long-short term memory). The order of VAR and VARMA means the number of preceding data that are used to predict the value at the present time.} 
\ree{The RNN-type predictive model consists of an \textit{encoder network} (two dense connected layers with only the first layer followed by tanh activation function) which takes the past observation signals and input signals as inputs to infer the initial hidden states, an \textit{RNN-type network} that recurrently generates the next $T_p=10$ steps of hidden states, and a \textit{decoder network} (two dense connected layers, and only the first layer followed with tanh activation function) converts the 10 steps of hidden states back to the original state space. 
The hyperparameters of the RNN-type model are tuned by using $MSE$ and $R^2$. The tuning results are summarized in Appendix Fig.~\ref{fig:JR_Model_Selection} and Fig.~\ref{fig:Epi_Model_Selection}. }

\ree{We use five metrics to quantify $T_p$-step prediction performance, including the Mean-Squared-Error ($MSE$), the Mean-Absolute-Error ($MAE$), the R-Square ($\mathrm{R^2}$), Median-Absolute-Error ($MeAE$) and Explained-Variance ($EV$).} 
Here, \ree{$MSE$ and $MAE$ quantify the mean of the sum of squared error (with higher penalty for larger error) and the mean of the sum of absolute error (with equal weight for different errors) between the predicted data and the ground truth, respectively. $MeAE$ quantifies the prediction robustness out of outliers. $EV$ and $ R^2$ compute the explained variance that accounts for and does not account for the systematic offset in the prediction respectively. The lower $MSE$, $MAE$ and $MeAE$ (indicated with $\downarrow$) and the higher $R^2$ and $EV$ (indicated with $\uparrow$), the better the seizure prediction.}

\ree{For seizure suppression, we compare the Koopman-MPC framework with the RNN-MPC framework (with GRU-cell and LSTM-cell). In fact, the Koopman model and the GRU-cell and LSTM-cell predictive models are the top 3 models with the best prediction performance. We evaluate the \textit{running time} in each MPC optimization process under the same hardware configuration (Intel Core i5-7600K CPU with a frequency of 3.80 GHz and a RAM of 16.00 GB). The running time is the time consumed in solving the quadratic programming problem when updating the optimal control signal. } 

\subsection{Hyperparameter Selection in Koopman model}

\begin{figure}[hpbt]
\centering
\includegraphics[width=0.9\linewidth]{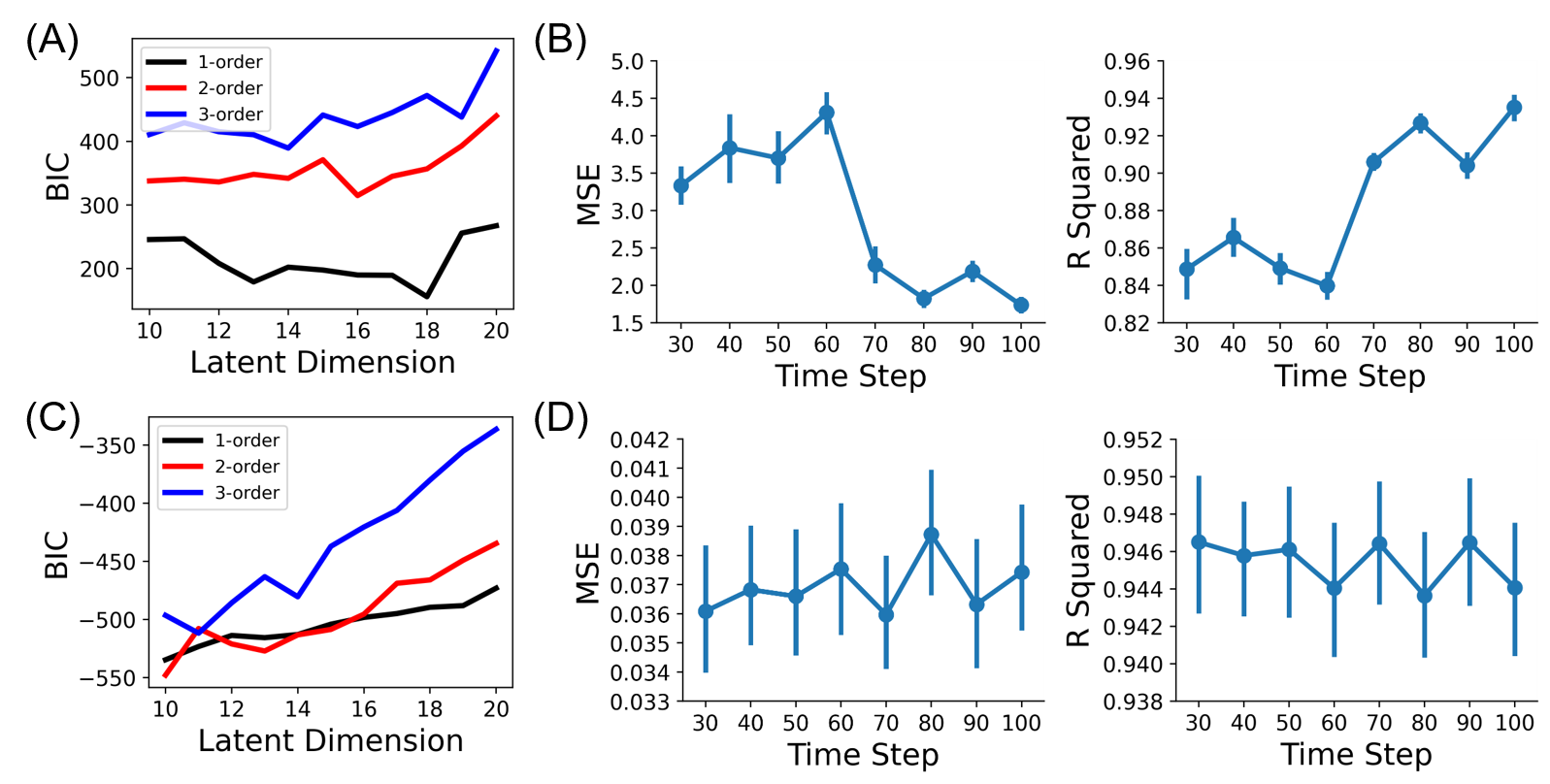}
\caption{\ree{Hyperparameter tunning in training Koopman Model for Jansen-Rit Model (A,B) and Epileptor Model (C,D). (A and C) The Bayesian Information Criterion (BIC) values for different dimensions and orders in the invariant subspace of the Koopman operator, (B and D) The mean squared error and explained variance given different time lengths of states in updating the Koopman operator. }}
\label{fig:Model_Selection}
\centering
\end{figure}

\ree{The dimension of the approximated Koopman operator is a hyperparameter in our study. The larger the dimension of the Koopman operator, the more complex the approximated linear dynamical system. Here, we introduce the Bayesian Information Criterion (BIC) for hyperparameter tuning. The definition of BIC is as follows,
\begin{equation}
    \label{Eq:Quantification}
    \mathrm{BIC} = n*LL+k*\log(n),
\end{equation}
where $n$ is the number of samples in the training dataset; $LL$ is the log-likelihood (\eg log of the mean squared error); and $k$ is the number of parameters in the model. In hyperparameter tuning of our model, we treat $n$ as the batch size of the training set, $LL$ as the log of $MSE$ in model prediction, and $k$ as the dimension of the Koopman Operator (the latent dimension in the latent layer multiplies the order of the linear system). The tuning of the dimension of the Koopman operator in two synthetic models is shown in Fig.~\ref{fig:Model_Selection} (A) and (C) respectively. }

\ree{After selecting the dimension of the approximated Koopman operator, we then tune the time lengths of states for updating the Koopman operator. We quantify the mean squared error and R squared given different time lengths of states. The results are shown in Fig.~\ref{fig:Model_Selection} (B) and (D) respectively. In summary, the combination of the optimal hyperparameter (latent dimension, order and time length) in training the Koopman model in the Jansen-Rit Model is latent dimension=18, order=1 and time length=100 and in Epileptor Model is latent dimension=10, order=2 and time length=70, respectively.}

\subsection{Prediction of Seizure Dynamics} 
\label{sec:results-Prediction}

\ree{The Koopman operator model and the RNN-type predictive models are trained to predict seizure dynamics. The VAR, VARMA and Kernel-based predictive models are online learning methods that do not require a training procedure.
In the training procedure, the network parameters (\eg the encoder and the decoder), the Koopman operator, and the control matrix of the Koopman operator model are learned, and the network parameters and the RNN-cell parameters of RNN-type predictive models are learned. In the test procedure, the Koopman operator and the control matrix can be updated in real-time explicitly thanks to its linear property.} 

\begin{figure}[thbp]
\centering
\includegraphics[width=\linewidth]{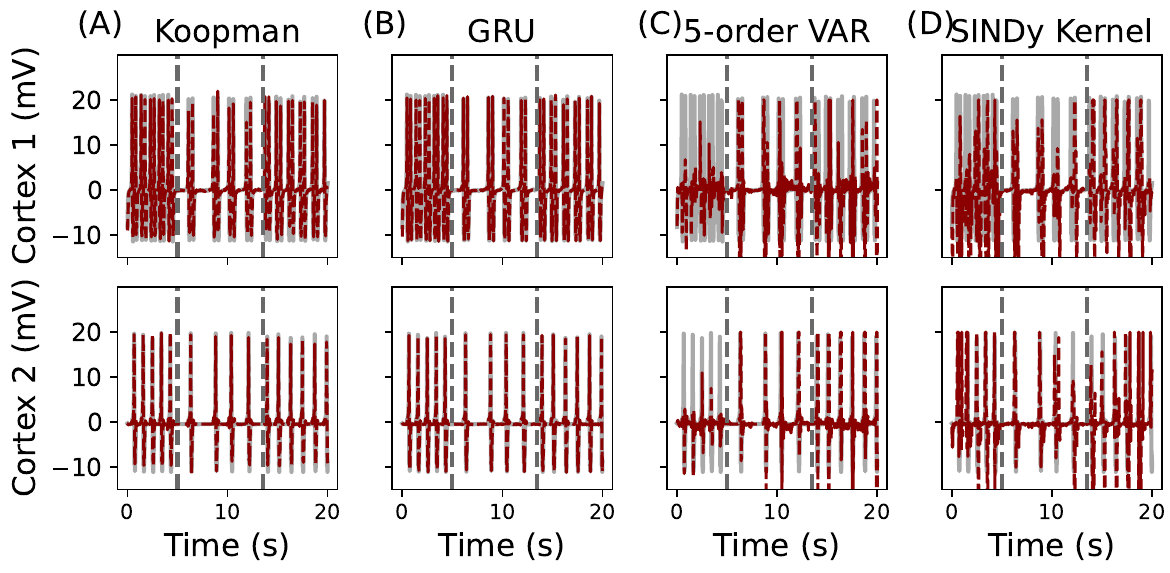}
\caption{\ree{The prediction of the synthetic seizure-like EEG waves generated by the Jansen-Rit model with two interconnected cortical columns. From (A) to (D) are the prediction result of (A) Koopman-operator, (B) GRU Model, (C) 5-order VAR  and (D) the Kernel based method (Polynomial kernel and Fourier Kernel). Note, the gray line is the ground truth time series, the red line is the prediction result. The middle part  is the interictal period (5-13s) and the others are the ictal period (0-5s and 13-20s).}}
\label{fig:two_column_synthetic_JR}
\centering
\end{figure}

\subsubsection{Prediction of the synthetic data}

\begin{figure}[thbp]
\centering
\includegraphics[width=\linewidth]{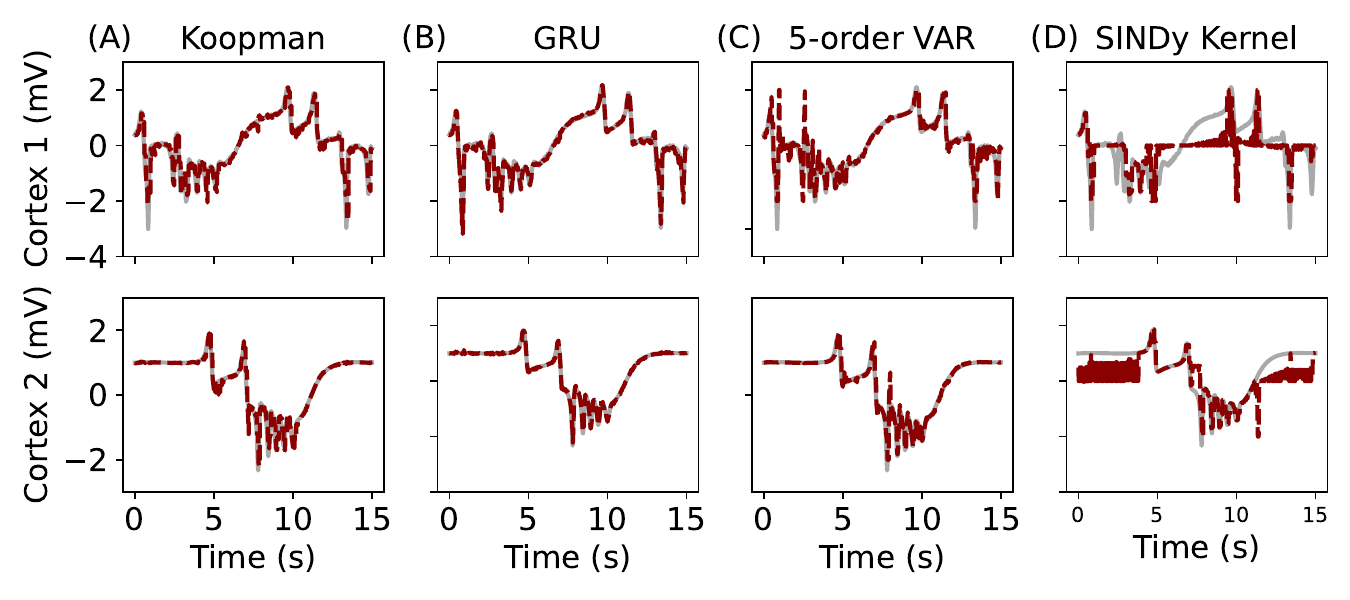}
\caption{The prediction of the synthetic seizure-like EEG waves generated by the Epileptor model with two interconnected cortical columns. From (A) to (D) are the prediction result of (A) Koopman-operator , (B) GRU Model, (C) 5-order VAR  and (D) the Kernel based method (Polynomial kernel and Fourier Kernel). Note, the gray line is the ground truth time series, the red line is the prediction result. }
\label{fig:two_column_synthetic_Epi}
\centering
\end{figure}

\ree{Prediction results from the synthetic data with the coupled Jansen-Rit model and Epileptor model are shown in Fig.~\ref{fig:two_column_synthetic_JR} \&~\ref{fig:two_column_synthetic_Epi}, Table~\ref{table:fitting_synthetic_JR} \& \ref{table:fitting_synthetic_Epileptor} respectively.
Specifically, Fig.~\ref{fig:two_column_synthetic_JR} \&~\ref{fig:two_column_synthetic_Epi} illustrate the prediction of temporal evolution when using the Koopman operator based model, GRU-cell model, 5-order VAR and SINDy model. The results show two deep learning models (\ie Koopman operator based model and the GRU model) can better predict future dynamics.}
Table~\ref{table:fitting_synthetic_JR} shows \ree{six quantification metrics (\ie $\#$ of parameters, $MSE$, $MAE$, $MeAE$, $EV$, and $R^2$) for the coupled JR model, and Table~\ref{table:fitting_synthetic_Epileptor} for the Epileptor model. 
Table~\ref{table:fitting_synthetic_JR} \& \ref{table:fitting_synthetic_Epileptor} demonstrate that the Koopman model outperforms or approximates to the baseline models. Our results suggest that the deep Koopman model can learn the invariant subspace and predict the future time series.}

\begin{table*}[tb]
\centering
\caption{The prediction performance in the synthetic Data (Coupled JR Model).}
\begin{tabular}{@{}lcccccc}
\hline
Model & $\#$ of param. & $MSE\downarrow$  & $MAE\downarrow$ & $MeAE\downarrow$ & $EV\uparrow$ & $R^2\uparrow$ \\ \hline


VAR (5-order) & 0 & 31.556±7.436 & 2.842±0.509  & 0.885±0.370 & -1.191±0.821 & -1.206±0.823\\
VAR (10-order) & 0 & 36.646±13.357 & 3.281±0.677  & 1.295±0.535 & -0.475±0.267 & -0.494±0.269\\
VARMA (5-order AR, 5-order MA) & 0 & 23.829±6.330  & 2.549±0.635  & 0.852±0.407 & -3.119±1.732 & -3.122±1.733\\

SINDy (Poly.+Four. Kernel) & 0 & 73.914±18.838 & 4.948±1.154 & 1.637±0.762 & -0.083±0.060 & -0.088±0.059  \\

Vanilla RNN-Cell (14-dim) & 4125 & 4.414±1.358 & 0.897±0.234  & 0.198±0.074 & 0.816±0.055 & 0.816±0.056\\
LSTM-Cell (20-dim)  & 6492 & 4.764±1.462 & 0.914±0.285 & \textbf{0.165±0.084} & 0.801±0.048  & 0.801±0.048\\
GRU-Cell (18-dim)  & 5231 & 4.212±1.322 & 0.880±0.248  & 0.174±0.085 & 0.819±0.027& 0.819±0.027\\
Koopman (18-dim,1-order) & 1460 & \textbf{1.578±0.796} & \textbf{0.559±0.132} & 0.170±0.067 & \textbf{0.930±0.059} & \textbf{0.930±0.060}\\
\hline
\end{tabular}
\label{table:fitting_synthetic_JR}
\end{table*}

\begin{table*}[htbp]
\centering
\caption{The prediction performance in the synthetic Data (the Epileptor Model).}
\begin{tabular}{@{}lcccccc}
\hline
Model & $\#$ of param. & $MSE\downarrow$  & $MAE\downarrow$ & $MeAE\downarrow$ & $EV\uparrow$ & $R^2\uparrow$ \\ \hline

VAR (5-order) & 0 & 0.053±0.021 & 0.101±0.021 & 0.035±0.010 & 0.923±0.039 & 0.923±0.039\\
VAR (10-order) & 0 & 0.059±0.023 & 0.108±0.022 & 0.038±0.011 & 0.917±0.040 & 0.916±0.040 \\

VARMA (5-order AR, 5-order MA) & 0 & 0.102±0.044 & 0.154±0.030 & 0.074±0.014 & 0.860±0.057 & 0.858±0.058\\

SINDy (Poly.+Four. Kernel) & 0 & 0.375±0.092 & 0.412±0.088 & 0.296±0.118 & 0.469±0.217 & 0.410±0.275\\

Vanilla RNN-Cell (10-dim \& 10-step)  & 1007 & 0.023±0.007 & 0.080±0.010 & 0.048±0.007 & 0.969±0.013 & 0.967±0.013\\

Vanilla RNN-Cell (18-dim)  & 5555 & 0.013±0.005 & 0.063±0.010 & 0.039±0.005 & 0.984±0.009 & 0.981±0.009\\

GRU-Cell (10-dim \& 10-step)  & 1267 & 0.017±0.006 & 0.065±0.010 & 0.035±0.005 & 0.976±0.011 & 0.976±0.012\\

GRU-Cell (20-dim \& 60-step)  & 9632 & \textbf{0.010±0.004} & \textbf{0.055±0.009} & \textbf{0.033±0.005} & \textbf{0.986±0.006} & \textbf{0.986±0.006}\\

LSTM-Cell (10-dim \& 10-step)  & 1397 & 0.017±0.006 & 0.067±0.010 & 0.037±0.006 & 0.976±0.011  & 0.975±0.011\\

LSTM-Cell (20-dim \& 60-step)  & 10092 & \textbf{0.010±0.004} & 0.055±0.010 & 0.033±0.006 & 0.986±0.007  & 0.985±0.007\\

Koopman (10-dim,2-order) & 712 & 0.035±0.010 & 0.090±0.013 & 0.038±0.006 & 0.949±0.021 & 0.949±0.021\\
\hline
\end{tabular}
\label{table:fitting_synthetic_Epileptor}
\end{table*}

\subsubsection{Prediction of the real iEEG data}

\begin{figure}[ht]
    \centering
    \includegraphics[width=\linewidth]{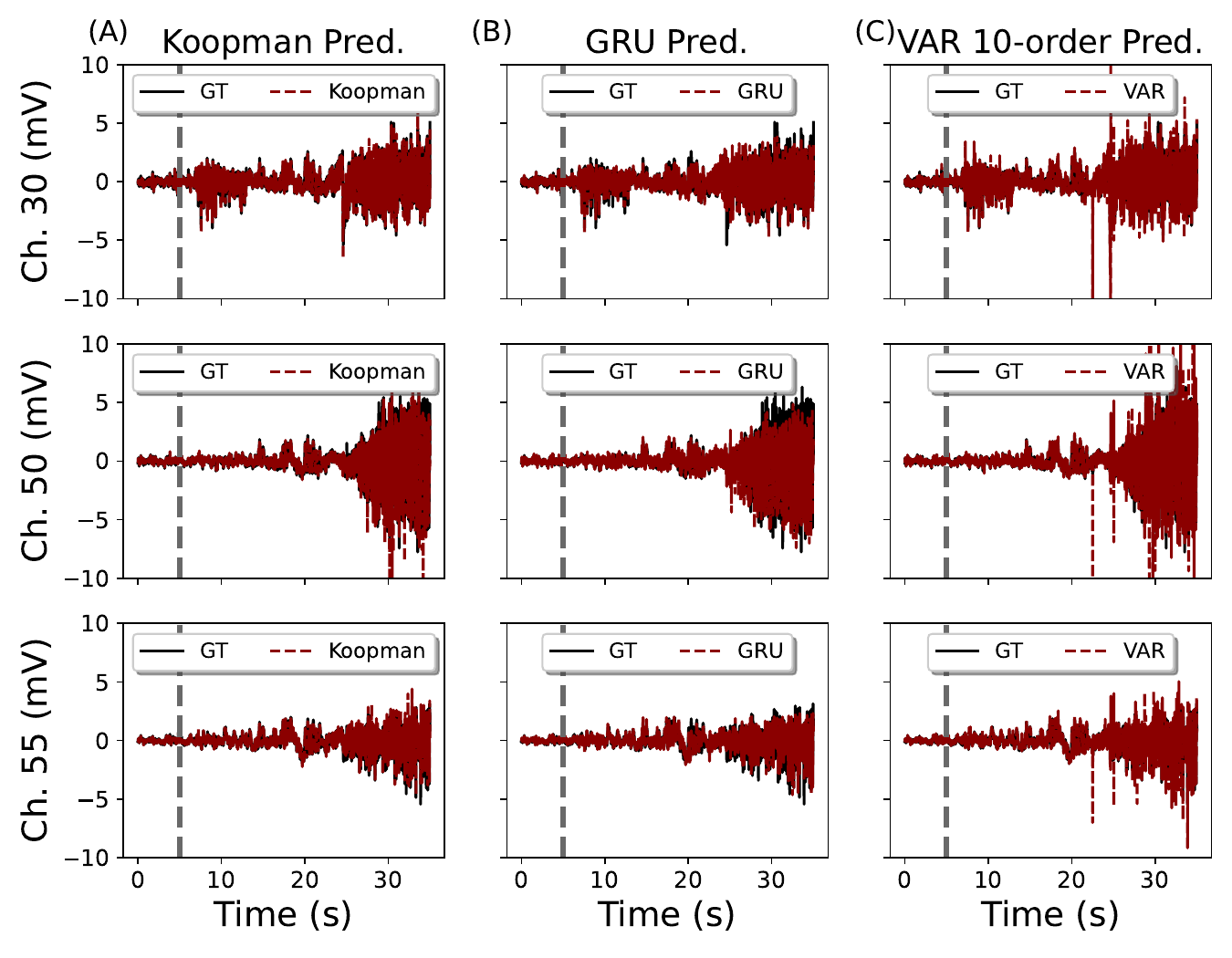}
    \caption{The visualization of prediction of testing real signals from interictal to ictal period in the second recording (randomly visualize 3 of 84 iEEG channels). (A) Koopman-based model; (B) GRU-based model; (C) 10-order vector autoregressive model. Note, the gray line is the time of seizure onset. GT is the ground truth times series.}
    \label{fig:real_data_prediction}
\end{figure}

\ree{We spilled the real 84-channel iEEG dataset into the training data and testing data. All the deep learning models are trained with the training data and tested with the testing data.} 
\ree{Fig.~\ref{fig:real_data_prediction} visualizes the prediction of seizure dynamics from the real iEEG \textbf{test} data using Koopman-based model, compared with the 10-order VAR model and a GRU-based model. 
The quantification metrics on the training and testing data are shown in Table~\ref{table:fitting_real}. Table~\ref{table:fitting_real} shows that although the RNN-type models (RNN-Cell, LSTM-Cell and GRU-Cell model) have better performance in training data, the performance is largely reduced in testing data. Online learning models (5-order VAR model, 10-order VAR model and Koopman model) have better generalizability.}



Results from synthetic data and real data indicate that the deep Koopman operator model can learn a finite-dimensional invariant subspace to transform the original nonlinear dynamical system to a linear system. 
The dynamics of the entire system are embedded into the approximated Koopman operator $\hat{\emph{\textbf{K}}}$ and the control matrix $\hat{\emph{\textbf{B}}}$, which can be re-calibrated at each step, underlying the flexibility of online learning. \ree{In contrast, the RNN-type models learn the hidden nonlinear dynamics and evolve under a pre-trained dynamical law. Its system properties cannot be updated online.
The vector autoregressive model fails to capture the highly nonlinear dynamics with large amplitudes in the predictions.} 

\begin{table*}[htbp]
\centering
\caption{The prediction performance in the real seizure dynamics Data (Epilepsy-iEEG-Multicenter-Dataset).}
\begin{tabular}{@{}lccccccccccc}
\hline
\multirow{2}{*}{Model} & \multirow{2}{*}{$\#$ of param.}  & \multicolumn{2}{c}{$MSE\downarrow$} & \multicolumn{2}{c}{$MAE\downarrow$} & \multicolumn{2}{c}{$MeAE\downarrow$} & \multicolumn{2}{c}{$EV\uparrow$} & \multicolumn{2}{c}{$R^2\uparrow$}\\ \cline{3-12} 
      &   & training  & test  & training  & test & training  & test & training  & test & training  & test  \\ \hline
VAR (5-order) & 0 & 0.084 & 0.111 & 0.108 & 0.125 & 0.051 & 0.060 & 0.807 & 0.812 & 0.807 & 0.812  \\
VAR (10-order) & 0 & 0.073 & 0.100 & \textbf{0.104} & \textbf{0.122} & \textbf{0.050} & \textbf{0.060} & 0.829 & 0.829 & 0.829 & 0.828  \\
SINDy (Poly.+Four. Kernel) & 0 & 0.961 & 0.235 & 0.329 & 0.229 & 0.164 & 157 & -6.437 & -1.687 & -6.474 & -1.693  \\
Vanilla RNN-Cell (100-dim) & 379k& 0.054 & 0.150 & 0.151 & 0.232 & 0.109 & 0.154 & 0.737 & 0.624 & 0.735 & 0.622\\
LSTM-Cell (100-dim)  & 410k & \textbf{0.028} & 0.127 & 0.111 & 0.207 & 0.081 & 0.132 & 0.850 & 0.691 & 0.850 & 0.690\\
GRU-Cell (100-dim) & 400k & \textbf{0.028} & 0.125 & 0.109 & 0.203 & 0.079 & 0.129 & \textbf{0.857} & 0.700 & \textbf{0.856} & 0.699\\
Koopman (140-dim, 1-order) & 102k & 0.049 & \textbf{0.072} & 0.112 & 0.139 & 0.070 & 0.086 & 0.851 & \textbf{0.858} & 0.851 & \textbf{0.858}\\
\hline
\end{tabular}
\label{table:fitting_real}
\end{table*}

\begin{figure}[thb]
\centering
\includegraphics[width=\linewidth]{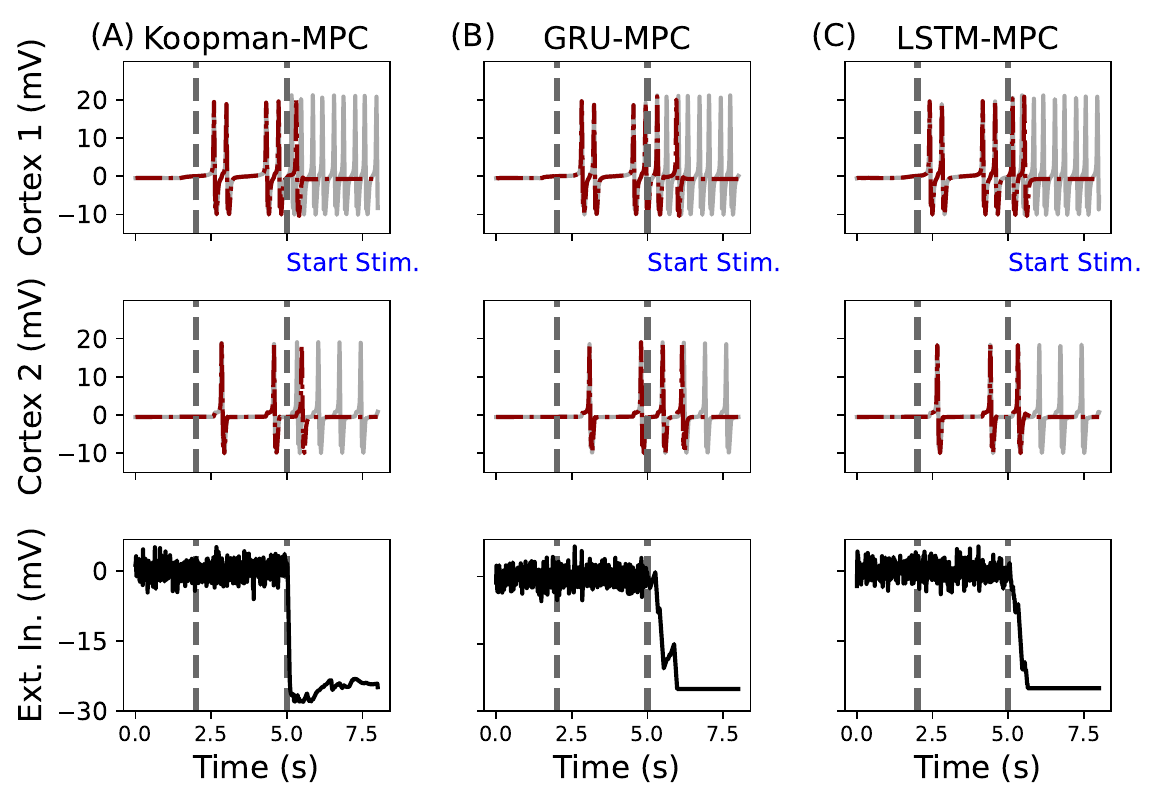}
\caption{\ree{Seizure suppression performance of MPC controllers with Jansen-Rit model. (A) Koopman-MPC controller; (B) GRU-MPC controller; (C) LSTM-MPC controller. The experiments are conducted on the simulation platform using coupled Jensen-Rit Model. From top to bottom, we show the EEG waveform in cortex 1, cortex 2, and the control signal applied to cortex 1, respectively. The red/gray line indicates the waveform with/without control. }}
\label{fig:suppressed_JR}
\end{figure}

\subsection{MPC-based Seizure Suppression}
\label{sec:results-MPC}
We test the seizure suppression performance in \ree{the Jansen-Rit model and the Epileptor model}. We examine the control ability and the computational efficiency of MPC optimization in seizure suppression \ree{with the Koopman-based model, the GRU-based model and the LSTM-based model.} Since seizure dynamics are highly nonlinear, it is difficult for a predictive model to track the dynamics of seizures over a long time. The Koopman-MPC model therefore periodically re-estimates the Koopman operator $\hat{\emph{\textbf{K}}}$ and control matrix $\hat{\emph{\textbf{B}}}$. 
To be noted, the network parameters in the Koopman-based model and \ree{RNN-type models (LSTM-Cell and GRU-Cell model) are not updated once the models have been trained.} Only the linear dynamical system parameters (\ie $\hat{\emph{\textbf{K}}}$ and $\hat{\emph{\textbf{B}}}$) can be updated online. The optimal neurostimulation strategy 
under the MPC framework is solved by the linear quadratic programming solver.


\begin{figure}[thb]
\centering
\includegraphics[width=\linewidth]{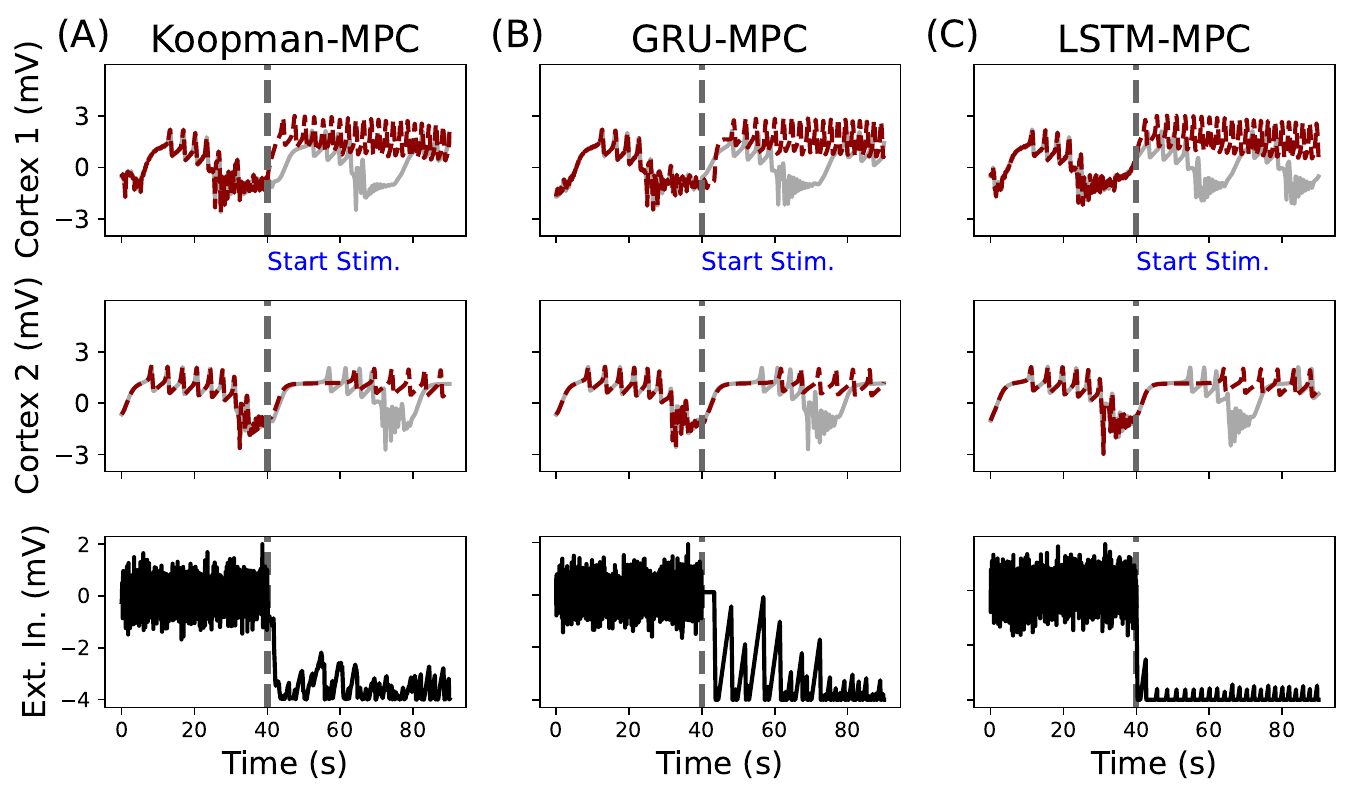}
\caption{\ree{Seizure suppression performance of MPC controllers with the Epileptor model. (A) Koopman-MPC controller; (B) GRU-MPC controller; (C) LSTM-MPC controller. The experiments are conducted on the simulation platform using a coupled Epileptor Model. From top to bottom, we show the EEG waveform in cortex 1, cortex 2, and the control signal applied to cortex 1, respectively. The red/gray line indicates the waveform with/without control. }}
\label{fig:suppressed_Epi}
\end{figure}


In \ree{LSTM-MPC and GRU-MPC frameworks, the hidden states lie in a highly nonlinear manifold space, rather than the linear transition space in the Koopman-MPC framework (Eq.~\eqref{Eq:MPC}). Thus, the nonlinear dynamical system in RNN-type models (\eg LSTM-Cell and GRU-Cell model) constructs a highly nonlinear quadratic programming optimization problem in MPC, which cannot be solved simply by a quadratic programming solver to obtain the optimal solution.}
To this end, we use a deep learning solution with gradient descent for optimization. Specifically, we transfer the \ree{LSTM-MPC and GRU-MPC} optimization problem into a pyTorch framework and solve it with the Adam solver (with a fixed learning rate of 0.01 and 10-step iterations).

\begin{table}[thpb]
\centering
\caption{The Computational Efficiency of Numerical Optimization for seizure suppression}
\begin{tabular}{@{}llcc}
\hline
Virtual platform & Model &  latent space 
& Running time (s) \\ \hline
\multirow{3}{*}{JR model} &
GRU-MPC  & 18-dim & 0.114±0.021 \\
& LSTM-MPC  & 20-dim & 0.114±0.019 \\
& Koopman-MPC & 18-dim  & \textbf{0.022±0.009} \\ \hline
\multirow{3}{*}{Epileptor model} & GRU-MPC  & 20-dim  & 0.113±0.007 \\
& LSTM-MPC  & 20-dim  & 0.115±0.005 \\
& Koopman-MPC & 10-dim  & \textbf{0.015±0.005} \\
\hline
\end{tabular}
\label{table:numerical_computation}
\end{table}

\ree{We conducted numerical experiments on the Jensen-Rit} \ree{simulation platform and the Epileptor platform to test the effects of the Koopman-MPC framework on seizure control. The stimulation signal is only applied to the epileptogenic foci (cortex 1).} We set a 10-step predictive horizon and a 10-step control horizon to obtain the optimal control signal for seizure suppression. In each experiment, we first collect some input-output data pairs for updating the model parameters in the deep Koopman operator model, as well as initializing the hidden state of \ree{the GRU and LSTM model. Then, the optimal control signal is designed by MPC controller and inserted into the Jansen-Rit model and the Epileptor model respectively. 
Fig.~\ref{fig:suppressed_JR} and Fig.~\ref{fig:suppressed_Epi} show the suppressed seizure signals (top and middle) and the corresponding control signals (bottom), which are provided by Koopman-MPC framework (Fig.~\ref{fig:suppressed_JR}A \& Fig.~\ref{fig:suppressed_Epi}A), GRU-MPC framework (Fig.~\ref{fig:suppressed_JR}B \& Fig.~\ref{fig:suppressed_Epi}B) and LSTM-MPC framework (Fig.~\ref{fig:suppressed_JR}C \& Fig.~\ref{fig:suppressed_Epi}C). }
In Fig.~\ref{fig:suppressed_JR}, although all models successfully suppress the seizures, the Koopman-MPC controller responds faster \ree{and with fewer seizure waves after triggering the stimulator. In Fig.~\ref{fig:suppressed_Epi}, all three frameworks can suppress the transition into the ictal period (the fast oscillation with negative amplitude). But suffering in the oscillation in the interictal period (the oscillation with positive amplitude). The main reason might be related to the stimulation artifacts.
The performance is further confirmed by the quantification of the computational efficiency in Table~\ref{table:numerical_computation}. The Koopman operator MPC framework only takes $0.022$ second in the Jansen-Rit model and takes $0.015$ second in the Epileptor model to obtain the optimal control signal at each time step}, which is sufficiently fast for the closed-loop neurostimulation applications.



\subsection{Online Learning of Koopman Operator for closed-loop neurostimualtion}
\label{online_learning_seizure_suppression}

\begin{figure}[ht]
    \centering
    \includegraphics[width=0.8\linewidth]{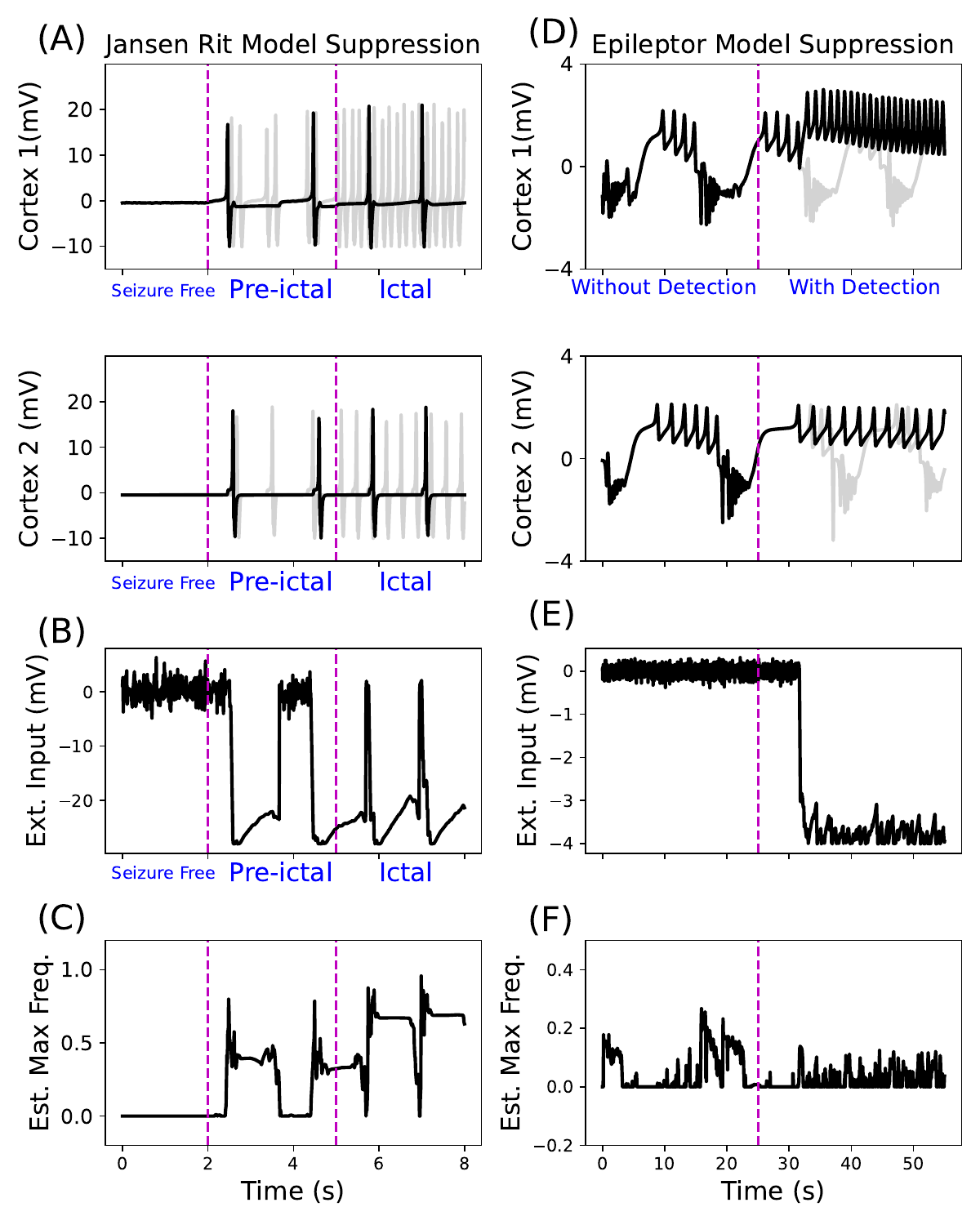}
    \caption{\ree{Online learning and Seizure Detection for closed-loop neuromodulation using Koopman-MPC framework in JR model (A,B,C) and Epileptor model (D,E,F) respectively.  (A,D) the EEG waveform at cortex 1 and 2 are presented. The black/gray line indicates the waveform with/without control. (B,D) The external input signal optimized by Koopman-MPC controller; (C,F) The maximum frequency of the estimated Koopman operator $\hat{\emph{\textbf{K}}}$.}}
    \label{fig:nonseizure_seizure_suppression}
\end{figure}

\ree{We further test whether the online learning Koopman-MPC framework can predict and control seizures during the state transition process in the Jansen-Rit model (\eg the state transition among seizure-free, preictal, and ictal period) and in the Epileptor model (\eg the state transition between interictal and ictal period).
In Jansen-Rit model, we manipulate the state transitions by varying the excitatory gain parameter $A$: from the seizure-free state, to the preictal period (slow oscillation), then to the seizure state (fast oscillation).
Specifically, EEG signals are in the non-ictal state for the first 2 seconds ($A=7.0$), then shift to the preictal period for 3 seconds ($A=7.2$), and eventually turn to the ictal state for the last 3 seconds ($A=7.8$). The Epileptor model simulates the interictal and ictal periods with three different time scales. The fastest scale reflects the fast oscillations, while the lowest scale governs the transition between the interictal and ictal periods. The gray lines in Fig.~\ref{fig:nonseizure_seizure_suppression}(A\&D) show the synthetic EEG signals without seizure control.}



\ree{We conduct eigendecomposition on the estimated Koopman operator $\hat{\emph{\textbf{K}}}$ to measure the maximum frequency of the seizure dynamics. Fig.~\ref{fig:nonseizure_seizure_suppression}(C) and (F) show the maximum complex part of the eigenvalues of $\hat{\emph{\textbf{K}}}$ in spectral analysis for the data generated by the Jansen-Rit model and the Epileptor model, respectively}. \ree{We find that the maximum spectral of the system is largely increased when the seizure-like wave occurred in both synthetic data, suggesting that spectral analysis of the Koopman-based model might offer a potential biomarker of seizure onset and need to be validated with more real dataset.}


\ree{Inspired by the results of spectral analysis, we design a closed-loop neurostimulation system using the maximum frequency of $\hat{\emph{\textbf{K}}}$ (\ie $f$) as a biomarker. According to Fig.~\ref{fig:nonseizure_seizure_suppression}(C), we set a threshold $f_{\text{thres}}=0.1$. When $f>f_{\text{thres}}$, the controller is actuated and triggers the neurostimulation to suppress seizure-like waves.
We firstly conduct experiments to validate our closed-loop neurostimulation system in Jansen-Rit model. As shown in Fig.~\ref{fig:nonseizure_seizure_suppression} (A-C), the neurostimulation is triggered when $f>f_{\text{thres}}$. The seizures are immediately suppressed after the controller is actuated. When the Koopman-MPC framework detects the seizure dynamics, it triggers the stimulator again.} 

\ree{The closed-loop neurostimulation system is also tested in the Epileptor model (Fig.~\ref{fig:nonseizure_seizure_suppression}D-E). We set the threshold of the maximum frequency $f_{\text{thres}}=0.1$ and the threshold of EEG amplitude $V_{cortex}=0$ as biomarker in the Koopman-MPC framework. The detector starts detecting at 25s. Once triggering the neurostimulator, the seizure wave with fast oscillation and negative amplitude is controlled, as the black line (with Koopman-MPC control) vs the gray line (without control) in Fig.~\ref{fig:nonseizure_seizure_suppression}D. Despite considerable performance to suppress huge negative bursts, the small interictal spikes (\eg positive oscillations) cannot be completely suppressed. This may be related to the effect of stimulation artifacts or the interaction of internal states in the Epileptor model.} 


\section{Discussion}

In this work, we propose a novel deep Koopman-MPC framework for closed-loop electrical neurostimulation. A tailored autoencoder is employed to learn the invariant subspace of the Koopman operator, in which the coordinate transformation maps the nonlinear dynamics into finite-dimensional linear dynamics. The approximation of the Koopman operator provides sufficient prediction horizon for seizure prediction, which can be integrated into the MPC controller for seizure suppression.

\subsection{Prediction and Control for Seizure Dynamics}

The epileptic seizures can be modeled as a dynamical system~\cite{chang2020model,chatterjee2020fractional,jirsa2014epileptor}. System identification is an essential step for modeling and prediction of seizure dynamics. 
However, the accuracy and the complexity of the identified model is a trade-off in modeling seizure dynamics, especially when the model is integrated into the MPC framework for designing optimal neurostimulation in real-time.
In our work, the proposed Koopman operator based model has higher prediction accuracy compared with \ree{VAR model, Kernel-based model and nonlinear RNN-type models} (Table~\ref{table:fitting_synthetic_JR},~\ref{table:fitting_synthetic_Epileptor} and~\ref{table:fitting_real}), while maintaining a considerable size of network parameters. One possible reason is that the Koopman model learns an invariant subspace that captures sufficient nonlinear patterns in the approximated Koopman operator for seizure prediction~\cite{lusch2018deep}. A great merit of the Koopman-based model is its ability to online update the Koopman operator, whereas the \ree{RNN-type models only learn the dynamics of the training data, and they cannot update its governing function online.} Another merit of the Koopman model is its linear property, which greatly facilitates the convex optimization in MPC control and provides a unique optimal solution with high computational efficiency (Table~\ref{table:numerical_computation}). In contrast, the \ree{RNN-type MPC framework} contains a highly nonlinear optimization problem, which is typically computationally expensive. In fact, there is a trade-off between model complexity and computational efficiency when designing models for real-time closed-loop control. A more complex \ree{nonlinear} model might better identify the system dynamics, but it would suffer from the high computational cost in optimization. Importantly, the proposed Koopman-MPC framework, to some extent, breaks the speed-accuracy tradeoff by combining a finite-dimensional approximated Koopman operator model and a linear MPC optimal controller. 
\ree{Beyond the invasive neurostimulation investigated in this study, the Koopman-MPC framework might be extended to broader neurostimulation protocols involving other types of stimulation (e.g., transcranial magnetic stimulation (TMS) and transcranial direct current stimulation (tDCS) and Vagus nerve stimulation (VNS)) in the future.}

\subsection{Seizure simulation platform}
For ethical reasons, it is difficult to directly collect input-output data of patients with epilepsy. Therefore, using a computational model to generate neural dynamics is an alternative way \ree{to validate} control strategies~\cite{kajishima2017computational,hodgkin1952quantitative}. In this study, we applied \ree{the Jansen-Rit model and the Epileptor model} to validate the Koopman model for seizure prediction and to test the neurostimulation policy produced by the MPC controller. As the macroscopic cortical column-level neural mass model, the Jansen-Rit model accounts for \ree{the dynamical interactions among three populations of cells} (i.e., excitatory, inhibitory, and pyramidal neuron populations)~\cite{jansen1995electroencephalogram}. 
By varying the bifurcation parameter, the average excitatory synaptic gain $A$, the model can generate neural activity in healthy or seizure conditions~\cite{jansen1995electroencephalogram}. \ree{The Epileptor model is also a neural mass model that simulates the state transitions between the interictal and ictal period with three different time scales, the fast time scale controls the high-frequency oscillation of spike-wave while the slow time scale mimics the state transitions~\cite{jirsa2014epileptor}. }
Also, many other dynamical models have been proposed to simulate neurodynamics with physics laws for exploring the underlying neural mechanisms. For instance, the Kuramoto model, as a phase oscillator, is used to synthesize the fMRI phase dynamics series~\cite{fukushima2020structural} and to study the cognitive process~\cite{zheng2021kuramoto, bick2020understanding}. 
It would be of high interest to build a more realistic seizure simulation platform for testing different control policies and even for performing virtual surgery.

\subsection{Limitations and future works}
It is worthy to mention the limitations and future directions of our work. First, we did not collect input-output data from epilepsy patients in a biological experiment. Instead, we ran numerical experiments with the Jansen-Rit model \ree{and Epileptor model} to simulate the EEG data and conduct MPC control. 
Second, we did not study the controllability of the identified system, which can be characterized by the approximated Koopman operator and control matrix~\cite{han2020deep}. \ree{The controllability of a linear dynamical system reflects whether the system can reach any state in finite control steps.} Studying the controllability of neural dynamical systems will broadly advance neuroscience applications (\eg cognitive control, neurofeedback control). 
Lastly, we did not consider the effects of noise in our model. The EEG denoising techniques~\cite{zhang2021eegdenoisenet,yu2021embedding} will contribute to the robustness of Koopman-based models in real-world applications.

The brain is a complex coupled dynamical system~\cite{bullmore2012economy,liu2017detecting}. Modeling and controlling the brain network while balancing speed and accuracy is challenging. The development of neuromodulation techniques and cybernetics will promote each other. 
In our work, we only consider two cortical regions dynamics and their neuromodulation, controlling the complex brain network dynamics requires hierarchical network models and more advanced control theory~\cite{nakahira2021diversity,yuan2019data}. The development of control theory would provide new tools for neurostimulation, including the selection of optimal targeted stimulation regions, the robustness and safety of stimulation protocols.

\section*{Acknowledgment}

The authors gratefully acknowledge the anonymous reviewers for their insightful comments, Mr. Rongwei Liang for discussions and editing, Dr. Yi Yao and Prof. Haiyan Wu for their useful suggestions. All authors declare to have no conflict of interests.

\bibliographystyle{IEEEtran}
\bibliography{mybibliography}


\appendix




\subsection{The coupled Jansen-Rit Model}
\label{jansen_rit_model}
\ree{The Jansen-Rit model describes the dynamical interaction of three subpopulations, the pyramidal cells, the excitatory feedback interneurons and the inhibitory feedback interneurons. It can be extended into a coupled Jansen-Rit model to simulate neural activity in two interconnected cortical areas in Eq.\eqref{eq:double_jr_model}. Parameters of the coupled Jansen-Rit Model are summarized in Table~\ref{table:para-double_JR-model}.}
\begin{equation}
\begin{scriptsize}
    \begin{cases}
    \dot{y}_{0}=y_{3} \\
    
    \dot{y}_{3}=A a S(y_{1}-y_{2})-2 a y_{3}-a^{2} y_{0} \\

    \dot{y}_{1}=y_{4} + Input1\\
    \dot{y}_{4}=A a(p+C_{2} S(C_{1} y_{0})+K_{2} y_{13})-2 a y_{4}-a^{2} y_{1} \\

    \dot{y}_{2}=y_{5} \\
    
    \dot{y}_{5}=B b C_{4} S(C_{3} y_{0})-2 b y_{5}-b^{2} y_{2} \\

    \dot{y}_{6}=y_{9} \\
    
    \dot{y}_{9}=A^{\prime} a S(y_{7}-y_{8})-2 a y_{9}-a^{2} y_{6} \\

    \dot{y}_{7}=y_{10} \\

    \dot{y}_{10}=A^{\prime} a(p^{\prime}+C_{2}^{\prime} S(C_{1}^{\prime} y_{6})+K_{1} y_{12})-2 a y_{10}-a^{2} y_{7} \\

    \dot{y}_{8}=y_{11} \\

    \dot{y}_{11}=B^{\prime} b{C_{4}^{\prime} S(C_{3}^{\prime} y_{6})}-2 b y_{11}-b^{2} y_{8}(t) \\

    \dot{y}_{12}=y_{14} \\

    \dot{y}_{14}=A^{\prime} a_{\mathrm{d}} S(y_{1}-y_{2})-2 a_{\mathrm{d}} y_{14}-a^{2} y_{12} \\
    
    \dot{y}_{13}=y_{15} \\
    
    \dot{y}_{15}= A^{\prime} a_{\mathrm{d}} S(y_{7}-y_{8})-2 a_{\mathrm{d}} y_{15}-a^{2} y_{13}
    \end{cases}
\end{scriptsize}
\label{eq:double_jr_model}
\end{equation}

    
    
    
    

\ree{where the Sigmoid function $S(v) = \frac{2e_0}{1 + e^{r(v_0 - v)}}$, $Input1$ is the external input applied in cortex 1. }

\begin{table}[h]
\scriptsize
\centering 
\caption{Parameters of the coupled Jansen-Rit model}
\begin{tabular}{c|p{4cm}|c} 
\hline
Parameters & Description & Values \\
\hline
$A$ & Average excitatory synaptic gain & 7.8 mV  \\ 
$B$ & Average inhibitory synaptic gain & 22 mV \\ 
\hline
$C_1$ & Average synaptic connectivity & 135      \\
$C_2$ & Average synaptic connectivity & 108     \\
$C_3$ & Average synaptic connectivity & 33.75     \\
$C_4$ & Average synaptic connectivity & 33.75      \\
\hline
$A^{\prime}$ & Average excitatory synaptic gain & 7 mV  \\ 
$B^{\prime}$ & Average inhibitory synaptic gain & 22 mV \\ 
\hline
$C_1^{\prime}$ & Average synaptic connectivity & 135      \\
$C_2^{\prime}$ & Average synaptic connectivity & 108     \\
$C_3^{\prime}$ & Average synaptic connectivity & 33.75     \\
$C_4^{\prime}$ & Average synaptic connectivity & 33.75      \\
\hline
$a_d$ & delay parameter & $a/3$      \\
\hline
$a$ & Average synaptic time delay for excitatory population & 100 Hz \\ 
$b$ & Average synaptic time delay for inhibitory population & 50 Hz  \\
\hline
$v_0$ & the postsynaptic potential for which a 50\% firing rate is achieved &  6 mV     \\
$r$ & Steepness of Sigmoid function & 0.56 m/V      \\
$e_0$ & maximum firing rate of the neural
population & 2.5 Hz      \\
\hline
$K_1$ & Connectivity constants & 100      \\
$K_2$ & Connectivity constants & 100      \\
\hline
\end{tabular}
\label{table:para-double_JR-model}
\end{table}

    


    

    






    
    

\subsection{Epileptor Model}
\label{epileptor_description}
\ree{The parametrized epileptor neural field model is composed of five state variables in three different time scales: fast, slow and intermediate time scales. The epileptor model can be summarized as }
\begin{equation}
\scriptsize
\label{eq:epileptor}
\begin{aligned}
\dot{x}_{1,i}=& \frac{1}{\tau_{1}}[(y_{1,i}-f_{1}\left(x_{1,i}, x_{2,i}\right)-z_{i}+3.1)+Input_{i}] \\
\dot{y}_{1,i}=& \frac{1}{\tau_{1}}[1-5 x_{1,i}^{2}-y_{1,i}] \\
\dot{z}_{i}=& \frac{1}{\tau_{0}}\left[4\left(x_{1,i}-x_{0}\right)-z_{i}- \sum\limits_{j = 1}^N {{K_{i,j}}}  \cdot ({{x_{1,j}} - {x_{1,i}}})\right] \\
\dot{x}_{2,i}=&\frac{1}{\tau_{1}}[-y_{2,i}+x_{2,i}-x_{2,i}^{3}+0.45+0.002 g\left(x_{1,i}\right) \\
&-0.3(z_{i}-3.5)] \\
\dot{y}_{2,i}=& \frac{1}{\tau_{2}}\left[-y_{2,i}+f_{2}\left(x_{2,i}\right)\right]
\end{aligned}
\end{equation}

\ree{where}

\begin{equation}
\begin{scriptsize}
\begin{aligned}
f_{1}\left(x_{1}, x_{2}\right) &=\left\{\begin{array}{ll}
x_{1}^{3}-3 x_{1}^{2} & x_{1}<0, \\
{\left[x_{2}-m-0.6(z-4)^{2}\right] x_{1}} & x_{1} \geq 0,
\end{array}\right.\\
f_{2}\left(x_{2}\right) &=\left\{\begin{array}{ll}
0 & x_{2}<-0.25, \\
6\left(x_{2}+0.25\right) & x_{2} \geq-0.25,
\end{array}\right.\\
g\left(x_{1}\right) &=\int_{t_{0}}^{t} e^{-\gamma(t-\tau)} x_{1}(\tau) d \tau
\end{aligned}  
\end{scriptsize}
\end{equation}

\ree{Here we set $\tau_{0}=10, \tau_{1}=0.05, \tau_{2}=2.5$. $x_0$ is the excitability parameter that regulates the epileptogenic cortex. $x_0=-1.4$ in cortex 1 (Seizure onset zone) and $x_0=-2.2$ in cortex 2 (Propagation Zone). $Input_{i}$ is the external input inserted at cortex $i$. We only insert external input to cortex 1.}

\subsection{RNN-type Model for System Identification} \label{app:SI}
\begin{figure}[htpb]
\centering
\includegraphics[width=\linewidth]{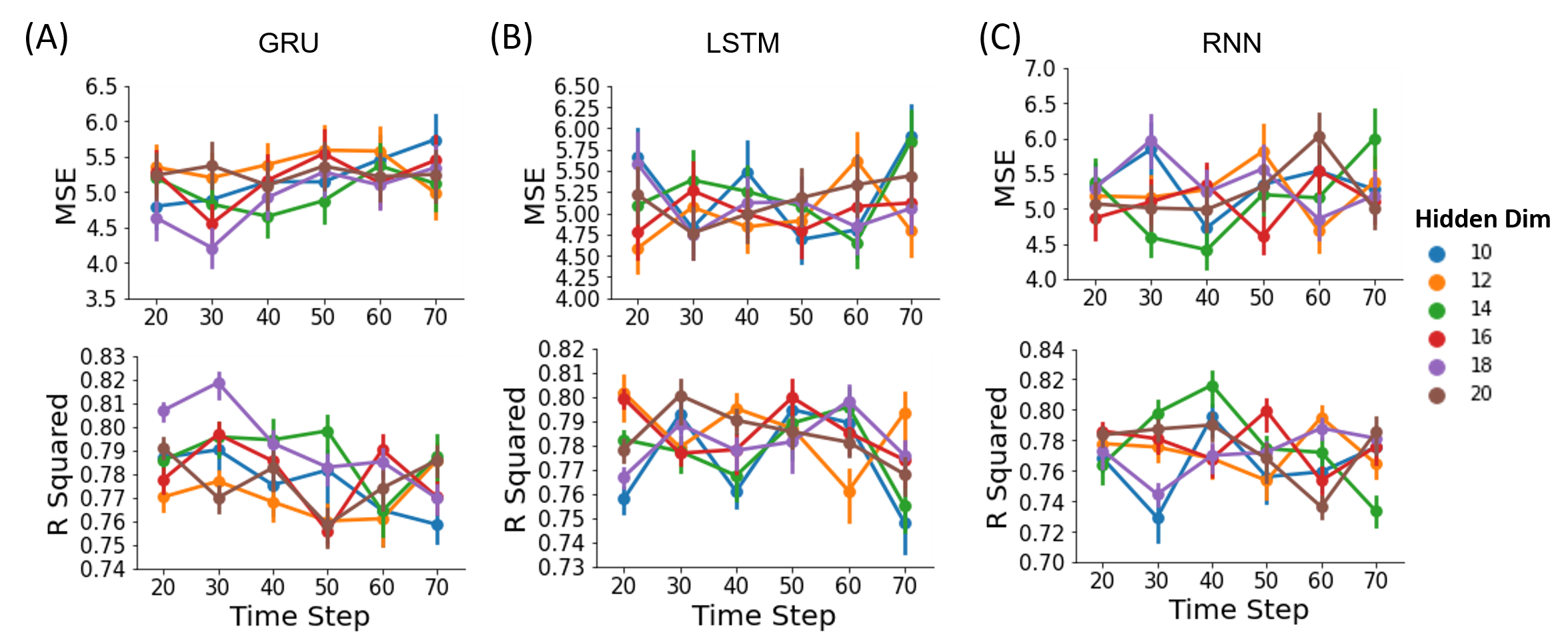}
\caption{\ree{Hyperparameter tunning in training RNN-type model in Jansen-Rit Model. The quantification results about mean squared error (top) and R squared (bottom) of (A) GRU-Cell model, (B) LSTM-Cell model, (C) RNN-Cell model.}}
\label{fig:JR_Model_Selection}
\centering
\end{figure}

\begin{figure}[htpb]
\centering
\includegraphics[width=\linewidth]{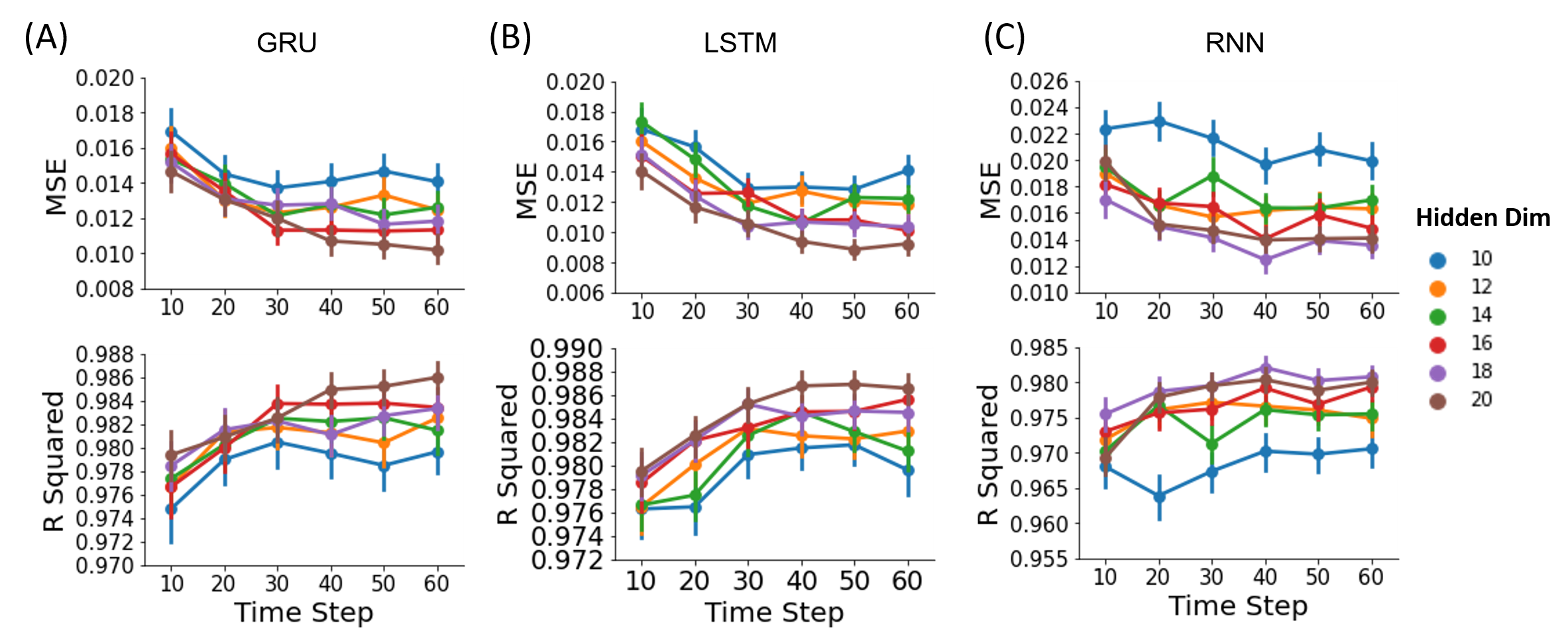}
\caption{\ree{Hyperparameter tunning in training RNN-type model in Epileptor Model. The quantification results about mean squared error (top) and R squared (bottom) of (A) GRU-Cell model, (B) LSTM-Cell model, (C) RNN-Cell model.}}
\label{fig:Epi_Model_Selection}
\centering
\end{figure}

\ree{The RNN-type models have been validated in fluid flow dynamics~\cite{bieker2020deep} and process industries~\cite{lanzetti2019recurrent}. Here, we apply it as baseline model to predict the seizure-like dynamics. We instantiate RNN-type predictive model with three types of recurrent cells (\ie RNN-Cell, LSTM-Cell and GRU-cell). The previous states and inputs are used to infer the initial hidden state of the recurrent cell, and the hidden dimension reflects model complexity. The dimension in the output layer of the encoder and the input of the decoder is determined by the hidden dimension of the RNN-type cell. The input dimension of the encoder is determined by number of state and time steps of previous state-input pair. 
We choose the hyperparameters of each RNN-type model with the lowest $MSE$ and the highest $R^2$(Fig.~\ref{fig:JR_Model_Selection} and ~\ref{fig:Epi_Model_Selection}). }

\end{document}